\documentclass[useAMS,usenatbib]{mn2e}

  \usepackage{graphicx}
   \usepackage{lscape}
 \usepackage{rotating}
  \usepackage{array}
  \usepackage{flafter}

\title{The Decomposed Bulge and Disk Size-Mass Relations of Massive Galaxies at $1<z<3$ in CANDELS }
\author[V. A. Bruce]
{V.A. Bruce$^{1}$\thanks{E-mail: vab@roe.ac.uk}, J.S. Dunlop$^{1}$, R.J. McLure$^{1}$, M. Cirasuolo$^{1,2}$, 
F. Buitrago$^{1}$,
\newauthor
R.A.A. Bowler$^{1}$, T.A. Targett$^{3}$, E.F. Bell$^{4}$, D.H. McIntosh$^{5}$, A. Dekel$^{6}$, 
\newauthor
S.M. Faber$^{7}$, H.C. Ferguson$^{8}$, N.A. Grogin$^{8}$, W. Hartley$^{9}$,  D.D. Kocevski$^{10}$,  
\newauthor
 A.M. Koekemoer$^{8}$, D.C. Koo$^{7}$, E.J. McGrath$^{11}$ \\
$^1$SUPA\thanks{Scottish Universities Physics Alliance} Institute for Astronomy, University of Edinburgh, Royal Observatory, Edinburgh EH9 3HJ\\
$^2$UK Astronomy Technology Centre, Science and Technology Facilities Council, Royal Observatory, Edinburgh EH9 3HJ\\  
$^3$Department of Physics and Astronomy, Sonoma State University,1801 East Cotati Avenue, Rohnert Park, CA 94928-3609, USA\\
$^4$Department of Astronomy, University of Michigan, Ann Arbor, MI 48109, USA\\
$^5$Department of Physics \& Astronomy, University of Missouri-Kansas City, 5110 Rockhill Road, Kansas City, MO 64110, USA\\
$^6$Racah Institute of Physics, The Hebrew University, Jerusalem 91904, Israel\\
$^7$UCO/Lick Observatory, Department of Astronomy and Astrophysics, University of California, Santa Cruz, CA 95064, USA\\
$^8$Space Telescope Science Institute, 3700 San Martin Drive, Baltimore, MD 21218, USA\\
$^9$ETH Z{\"u}rich, Institut f{\"u}r Astronomie, HIT J 11.3, Wolfgang-Pauli-Str. 27, 8093 Z{\"u}rich\\
$^{10}$Department of Physics and Astronomy, University of Kentucky, Lexington, KY 40506, USA\\
$^{11}$Department of Physics and Astronomy, Colby College, Waterville,ME 0490, USA }

\begin{document}

\date{}

\pagerange{\pageref{firstpage}--\pageref{lastpage}} \pubyear{2013}

\pagestyle{myheadings}
\markboth{V. A. Bruce et al.} {Decomposed Bulge and Disk Sizes at $1 < z < 3$}

\maketitle

\label{firstpage}

\vspace*{-1.5cm}

\begin{abstract}
We have constructed a mass-selected sample of $M_*>10^{11}\,{\rm M_{\odot}}$ galaxies at $1<z<3$ in the CANDELS UDS and COSMOS fields and have decomposed these systems into their separate bulge and disk components according to their $H_{160}$-band morphologies. By extending this analysis to multiple bands we have been able to conduct individual bulge and disk component SED fitting which has provided us with stellar-mass and star-formation rate estimates for the separate bulge and disk components. These have been combined with size measurements to explore the evolution of these massive high-redshift galaxies. By utilising the new decomposed stellar-mass estimates, we confirm that the bulge components display a stronger size evolution than the disks. This can be seen from both the fraction of bulge components which lie below the local relation and the median sizes of the bulge components, where the bulges are a median factor of $2.93\pm0.32$ times smaller than similarly massive local galaxies at $1<z<2$ and $3.41\pm0.58$ smaller at $2<z<3$; for the disks the corresponding factors are $1.65\pm0.14$ smaller at $1<z<2$ and $1.99\pm0.25$ smaller at $2<z<3$. Moreover, by splitting our sample into the passive and star-forming bulge and disk sub-populations and examining their sizes as a fraction of their present-day counter-parts, we find that  the star-forming and passive bulges are equally compact, star-forming disks are larger, while the passive disks have intermediate sizes. This trend is not evident when classifying galaxy morphology on the basis of single-S\'{e}rsic fits (ie. $n>2.5$ or $n<2.5$) and adopting the overall star-formation rates. Finally, by evolving the star-formation histories of the passive disks back to the redshifts when the passive disks were last active, we show that the passive and star-forming disks have consistent sizes at the relevant epoch. These trends need to be reproduced by any mechanisms which attempt to explain the morphological evolution of galaxies.

\end{abstract}


\newpage
\clearpage

\section{Introduction}
In recent years, increasingly detailed high-resolution morphological studies of massive galaxies at $z>1$ have provided strong evidence for evolution in the sizes of high-redshift galaxies, which are observed to be up to a factor of $\sim2-6$ ( e.g. \citealt{Daddi2005, Trujillo2006, Toft2007, Trujillo2007, Buitrago2008, Cimatti2008, Franx2008, vanDokkum2008, Damjanov2009} and \citealt{Cassata2010}) more compact than similarly massive present-day systems, with the most compact high-redshift systems also being seen to be the most  passive \citep{Toft2007, Kriek2009,Newman2012,McLure2013}.

Despite early suggestions that the sizes of these systems were under-predicted due to selection effects and measurement uncertainties in both mass and size \citep{Vanderwel2009,Muzzin2009,Mancini2010}, several spectroscopic campaigns \citep{Vanderwel2008,Newman2010, vandeSande2011, vandeSande2013, McLure2013} have subsequently provided more robust dynamical mass measurements. Additionally, tests of the sizes of simulated galaxies recovered by commonly adopted fitting procedures, such as GALFIT \citep{Peng2010galfit} and GALAPAGOS \citep{Barden2012}, by \citet{Haussler2007}, \citet{vanderWel2012}, \citet{Newman2012}, and more recently by \citet{Davari2014} for more complex morphological systems equivalent to local ellipticals, have shown that these size estimates are not significantly biased or under-estimated. These results, coupled with the deep morphological studies of small samples by \citet{Szomoru2010,Szomoru2012} and \citet{Trujillo2012} confirm the genuine compactness of high-redshift galaxies. 

However, in spite of mounting evidence for the evolution in the median size of the massive galaxy population with redshift, there remains debate over whether a significant fraction of the compact systems survive to the present day and, moreover, if the increase in the median size of galaxies is driven by the growth of individual systems or by the addition of newly quenched, larger, galaxies to the passive population. 

Recent studies such as those by \citet{Valentinuzzi2010a} and \citet{Poggianti2013} for cluster and field environments, respectively, have found that a significant fraction of local systems ($\sim 20\%$) are compact. These results, coupled with co-moving number density redshift evolution studies of compact, passive, galaxies \citep{Cassata2011, Carollo2013, Cassata2013,Poggianti2013b}, and the associated suggestions that the reported size evolution of massive galaxies may be over-estimated due to the effects of progenitor bias, have argued that the observed evolution in the median sizes of the massive galaxy population may not be primarily driven by the growth of individual systems (e.g. via minor merging or adiabatic expansion as proposed by \citealt{Khochfar2006, Naab2007, Hopkins2009, Shankar2011, Fan2008, Fan2010}), but instead by the addition of newly quenched, larger, galaxies to this population with time \citep{Carollo2013, Krogager2013,Poggianti2013b}. 

However, the newly reported prevalence of low redshift compact galaxies is in conflict with previous studies with SDSS, which found that as little as $\sim 0.03\%$ of the local population can be classified as compact \citep{Trujillo2009,Taylor2010}. Moreover, the latest co-moving number density study of passive (ETGs) over $0<z<3$ within the full CANDELS+3D-HST fields by \citet{vanderWel2014} shows that, whilst the overall co-moving number density of ``compact''  systems (defined simply as $r_{e}<2$\,kpc) does not appear to evolve strongly with redshift, the size distribution of galaxies within this ``compact''  classification does, such that the co-moving number density of small galaxies decreases with decreasing redshift. In addition to this, there is new evidence provided by the $z>1$ velocity dispersion study of \citet{Belli2014}, which reveals that by accounting for progenitor bias by considering systems at fixed velocity dispersions with redshift, the dominant contribution to the growth in sizes of passive galaxies between $0<z<2$ is the increase in the size of individual systems, rather than the addition of newly-quenched, larger galaxies. 

Existing studies of galaxy size evolution at $z>1$ have almost exclusively been conducted by fitting single-S\'{e}rsic light profiles to galaxies in order to measure their effective radii. However, it is becoming increasingly clear that within the $1<z<3$ regime massive galaxies are undergoing dramatic structural transformations from disk-dominated and visually disturbed morphological systems at $z>2$ to bulge dominated at lower redshifts \citep{Buitrago2011, Vanderwel2011, McLure2013, Mortlock2013, Mozena2013,
Wuyts2012, Bruce2012}. Therefore, in order to best conduct studies of the morphological evolution of galaxies at high redshift it is vital to trace both the bulge and disk components separately by decomposing galaxy morphologies into these two components. Previously, such bulge-disk decompositions have generally been conducted in the local Universe (e.g. \citealt{deJong1996, Allen2006, Cameron2009,Simard2011} and \citealt{Lackner2012}), where high-resolution imaging is more readily available. By contrast, bulge-disk decompositions at high redshifts have been
limited to small samples \citep{Vanderwel2011}. However, with the advent of large, high-resolution surveys such as CANDELS with {\it HST} WFC3 it is now possible to conduct the first decompositions at rest-frame wavelengths longer than the 4000\,$\rm\AA$ break (tracing the assembled stellar mass) for statistically significant mass-selected samples of high-redshift galaxies (\citealt{Bruce2012} and also \citealt{Lang2014} who conduct similar decompositions on stellar-mass maps). The results from our previous analysis \citep{Bruce2012} explicitly revealed that the bulge components display a much stronger size evolution with redshift than the disk components. However, this study was limited by the use of stellar-mass estimates determined for the entire galaxy which were sub-divided for the bulge and disk components based purely on the fraction of the $H_{160}$-band light which was attributed to each component. In the new study presented here we have further utilised the bulge-disk morphological decomposition approach to extend our analysis to the additional three photometric bands covered by CANDELS in order to conduct separate-component SED fitting. As described in Bruce et al. (2014a) this analysis has allowed separate stellar masses and, additionally, star-formation rates to be estimated for the individual bulge and disk components. Here, we use this new information to explore the size evolution of the bulge and disk components by separating them into star-forming and passive sub-populations based on their specific star-formation rates.

The structure of this paper is as follows. In Section 2 we provide a summary of our data-sets and the sample properties. This is followed in Section 3 by a brief overview of our decomposed multi-band morphological fitting and SED fitting procedure. In Section 4 we present the results from our analysis and explore the relation between single-S\'{e}rsic index fits and bulge-to-total light fractions, the decomposed size-mass relations of the galaxies in our sample and any trends with decomposed star-formation rates. We also use our decompositions to probe the fractional size evolutional of galaxies split into the decomposed disk and bulge, passive and star-forming sub-samples. These results are compared to those from existing studies in Section 5, after which we conclude with a discussion of our findings within the context of current galaxy size growth and quenching models. Finally, in Section 6 we summarise our main results.

Throughout this work we quote magnitudes in the AB system, and calculate all physical
quantities assuming a $\Lambda$CDM universe with $\Omega_{m}=0.3$, $\Omega_{\Lambda}=0.7$ and $H_{0}=70\rm{kms^{-1}Mpc^{-1}}$.

\section{Data}
We have used the high-resolution near-infrared {\it HST} WFC3/IR data from the CANDELS multi-cycle treasury programme \citep{Grogin2011, Koekemoer2011} centred on the UKIDSS Ultra Deep Survey (UDS; \citealt{Lawrence2007}) and the COSMOS \citep{Scoville2007,Koekemoer2007} fields. Both the CANDELS UDS and COSMOS near-infrared data comprise $4 \times 11$ WFC3/IR tiles covering a total area of 187\,arcmin$^2$ in each field, in both the F125W and F160W filters with 5-$\sigma$ point-source depths of 27.1 and 27.0 (AB mag) respectively. In addition to near-infrared data, we have also made use of the accompanying CANDELS {\it HST} ACS parallels in the F814W and F606W filters (hereafter $i_{814}$ and $v_{606}$).  The 5-$\sigma$ point-source depths are 28.4 for both the $i_{814}$ and $v_{606}$ bands in UDS and  28.5 in COSMOS. Approximately $80\%$ of the area of the UDS and COSMOS fields is covered by both ACS and WFC3 pointings.

\subsection{Supporting multi-wavelength data}
In addition to the near-infrared and optical imaging provided by {\it HST}, we have also utilised the multi-wavelength data-sets available in each field to constrain SED fitting and determine the physical properties for the galaxies in our sample. For the UDS these include: CFHT $u'$-band imaging; deep optical $B$, $V$, $R$, $i'$ and $z'$-band imaging  from the Subaru XMM-Newton Deep Survey (SXDS; \citealt{Sekiguchi2005}; \citealt{Furusawa2008}); $J$, $H$ and $K$-band UKIRT WFCAM imaging from Data Release 8 (DR8) of the UKIDSS UDS; and {\it Spitzer} 3.6\,$\mu$m, 4.5\,$\mu$m IRAC and 24\,$\mu$m MIPS imaging from the SpUDS legacy programme (PI Dunlop). For COSMOS they include: optical imaging in  $u'$, $g'$, $r'$, $i'$ and $z'$-bands from MegaCam CFHTLS-D2; $z'$-band from Subaru; $Y$,$J$,$H$ \& $Ks$ from Ultra-VISTA (PI Dunlop); and {\it Spitzer} 3.6\,$\mu$m, 4.5\,$\mu$m IRAC and 24\,$\mu$m MIPS imaging from the S-COSMOS survey (PI Sanders).
  
\subsection{Sample selection}
We have adopted the sample of Bruce et al. (2014a), which comprises a refined sub-sample from the $1<z_{phot}<3$ and $M_*>10^{11}\,{\rm M_{\odot}}$ CANDELS UDS sample of \citet{Bruce2012} (now making use of an updated stellar-mass fitting technique) and a similarly-selected sample in the CANDELS COSMOS field. Photometric redshifts were estimated using a code based on HYPERZ from \citet{Bolzonella2000}, following \citet{Cirasuolo2007} and were subsequently used to determine stellar-mass estimates. The stellar-mass estimates were based on the \citet{Bruzual2003} models with single-component exponentially-decaying star-formation histories with e-folding times in the range $0.3\leq \tau {\rm (Gyr)}\leq 5$ and with a minimum model age limit of 50 Myr. Our final sample contains 205 galaxies in the UDS and 191 galaxies in COSMOS with $1<z_{phot}<3$ and $M_*>10^{11}\,{\rm M_{\odot}}$. As the sample sizes and areas in the UDS and COSMOS fields are comparable and there is good agreement between the co-moving number densities of the two fields (Bruce et al. 2014a), in the following sections the science results are based on the combined UDS and COSMOS sample unless otherwise stated.

\subsection{Star-formation rates}
Finally, the star-formation rates for the UDS and COSMOS samples were estimated from the best-fit SED models and $24\mu$m fluxes by adopting the convention of \citet{Wuyts2011} where, if any of the objects in the sample have a $24\mu$m counterpart within a 2-arcsecond radius in the SpUDS and S-COSMOS catalogues, their star-formation rate is given by: $SFR_{UV+IR}[{\rm M_{\odot}yr^{-1}}]=1.09\times10^{-10}(L_{IR}+3.3L_{2800})/{\rm L_{\odot}} $
where $L_{2800}=\nu L_{\nu}(2800\rm\AA)$ and the contribution to $L_{IR}$ is taken over the wavelength range $8-1000\mu$m. For objects which do not have $24\mu$m counterparts, a value of:
$SFR_{UV, dust\,corrected}[{\rm M_{\odot}yr^{-1}}]=1.4\times10^{-28}{\rm L_{\nu}(ergs\,s^{-1}\,Hz^{-1}})$ is adopted \citep{Kennicutt1998}.

For completeness, we have also compared our distinction of passive and star-forming galaxies based on $sSFR= 10^{-10}{\rm yr}^{-1}$ to those from UVJ colour-cuts following \citet{Williams2009} and find that the two methods agree well.

\section{Multiple-Component Morphology Fitting}
Following Bruce et al. (2014a), the morphologies of the 396 objects in our combined sample have been fitted with both single and multiple-S\'{e}rsic light profiles using GALFIT \citep{Peng2010galfit}. This procedure makes use of an empirical PSF generated from a median stack of the brightest (unsaturated) stars in the individual fields and adopts a consistent object-by-object background determination, which has been calculated as the median value within an annular aperture centred on each source with an inner radius of 3\,arcsec and an outer radius of 5\,arcsec. GALFIT is then run on $6\times6$\,arcsec image stamps.

In addition to more basic single-S\'{e}rsic light-profile fits, we have also conducted a multiple-component S\'{e}rsic light-profile decomposition by fitting two sets of nested models to each object in our sample: 2 single-S\'{e}rsic models; n=free and n=free + PSF; and 6 multiple-component models; n=4 bulge, n=1 disk, n=4 + PSF, n=1 + PSF, n=4 + n=1 and n=4 + n=1 + PSF, where the PSF is included to account for any centrally concentrated light profile components such as nuclear starbursts or AGN. These multiple component models were run with a grid of different initial conditions to ensure that the fitting was robust against the $\chi^2$-minimisation routine becoming confined to local minima.

The best-fit multiple-component models within each of these nested sets were then determined by adopting the simplest model unless a more complex model fit was deemed statistically acceptable, as defined by: $\chi^{2}\le\nu+3\sqrt(2\nu)$, and if it satisfied $\chi^{2}_{complex}<\chi^{2}_{simple}-\Delta\chi^{2}(\nu_{complex}-\nu_{simple})$, where $\nu$ represents the number of degrees of freedom in the model (in effect the number of parameters), and $\Delta\chi^{2}$$(\nu_{complex}-\nu_{simple})$ is the 3-$\sigma$ value for the given difference in the degrees of freedom between the two competing fits. In addition to these criteria several additional constraints were applied to ensure that the best-fit models were physically realistic. This approach provided statistically acceptable multiple-component models for $\sim 85\%$ of the combined sample.

\subsection{Mock galaxy simulations}
In order to estimate the random and systematic uncertainties on our fitted morphological parameters we have conducted tests using simulated galaxies. Full details of the simulations and accompanying results are presented in Bruce et al. (2014a). In summary, we find that the we are able to recover $B/T$ ratios to within $10\%$ accuracy for $\sim80\%$ of objects, without any significant systematic bias. We also report that component sizes are robust to an accuracy of $10-20\%$, including systematic errors. However, we note that these are conservative estimates of uncertainties; disk components can be recovered more accurately and there is also a trend for models with smaller component sizes to be fitted with lower random and systematic uncertainties. It should also be noted that the accuracy with which we have been able to determine these fitted parameters relies heavily
on the high $S/N$ of the imaging data for our galaxy sample (typically $S/N>50$).

We have also utilised the mock galaxy simulations in Bruce et al. (2014a) to explore the effects of allowing a PSF component in the fitting. Here we discuss the cases where our best-fit  single-S\'{e}rsic fits have $n>10$ (with or without a fitted PSF component) in order to ascertain when GALFIT fits these un-physically high values. 

Out of 174 of these galaxies, we find that $80\pm9\%$ have one component with an effective radius of 1 pixel and the other with effective radius 20 pixels (the two extreme sizes modelled). By construction, only $12.5\%$ of {\it all} our models have this $r_{eff}=1+r_{eff}=20$ configuration. Within this subset  $64\pm4\%$ are from input models where the disk $r_{eff}=1$ and the remaining $36\pm4\%$ have bulge $r_{eff}=1$. Given the relatively small number statistics it is difficult to make robust statements, but it does not appear that these $n>10$ fits have any preferential $B/T$ light fraction, axis ratios, relative position angles or an increased probability of being fitted with a single-S\'{e}rsic+PSF model. 
Thus, it appears as though the unphysically high $n>10$ fits are a result of systems with large differentials in component sizes, which cannot be well-fit with a single S\'{e}rsic light profile.

Given that we retain all the single and multiple-component fits, it is also interesting to look at all the models which had an initial single-S\'{e}rsic $n>10$ fit, but where the {\it best-fit} then adopted a PSF component. Out of 312 of these initial single-S\'{e}rsic $n>10$ fits, $48\pm5\%$ (151 objects) were then best-fit by S\'{e}rsic+PSF, and only 13 objects with these S\'{e}rsic+PSF best-fits retained an $n>10$. 

Thus, our simulations confirm our assertion \citep{Bruce2012} that the adoption of S\'{e}rsic+PSF best-fit models are motivated by the inability of single-S\'{e}rsic fits to fully account for multiple components. The significance of the multiple-S\'{e}rsic+PSF best-fit models are discussed in detail in Bruce et al. (2014a).

\subsection{Extension to additional bands and decomposed SED fitting}

Having established $H_{160}$ bulge-disk decomposed morphological fits for all the objects in the combined UDS and COSMOS sample, in Bruce et al. (2104a) we were then able to extend this analysis to the other three bands available within CANDELS: $J_{125}$, $i_{814}$ and $v_{606}$ and conduct separate component SED fitting on the decomposed photometry (where the SED fits were further constrained at the extreme blue and red ends by the overall photometry for the objects) .
The full details of this procedure are presented in Bruce et al. (2014a) and here we highlight that this technique does not rely on the adoption of any functional forms to describe how the morphologies of these massive galaxies vary as a function of redshift, but instead, fixes all morphological parameters at the $H_{160}$-band best-fit values. This simplified approach accounts for colour gradients within the bulge+disk systems by allowing the bulge and disk component magnitudes to trade-off against each other and yields realistic colours for the bulge and disk components without any further constraints.

This decomposition technique provides several clear advantages. By providing individual stellar-mass and star-formation rates for the separate components (where burst and exponentially decaying star-formation history templates with $0.1\leq \tau {\rm (Gyr)}\leq 5$ can now be fitted due to the additional degrees of freedom included from the bulge-disk decomposition) we are able to: i) explore the fully decomposed bulge and disk size-mass relations; ii) study the trends with star-formation rate for the separate bulge and disk components, which has provided new insight into the links between quenching and size evolution; iii) given our $M_*>10^{11}\,{\rm M_{\odot}}$ sample selection, the decomposition also allows us to probe the lower mass envelopes of the individual components.

\section{Results}
\subsection{Correlation Between Single and Multiple-Component Model Morphologies}
\begin{figure*}
\begin{center}
\includegraphics[scale=0.9]{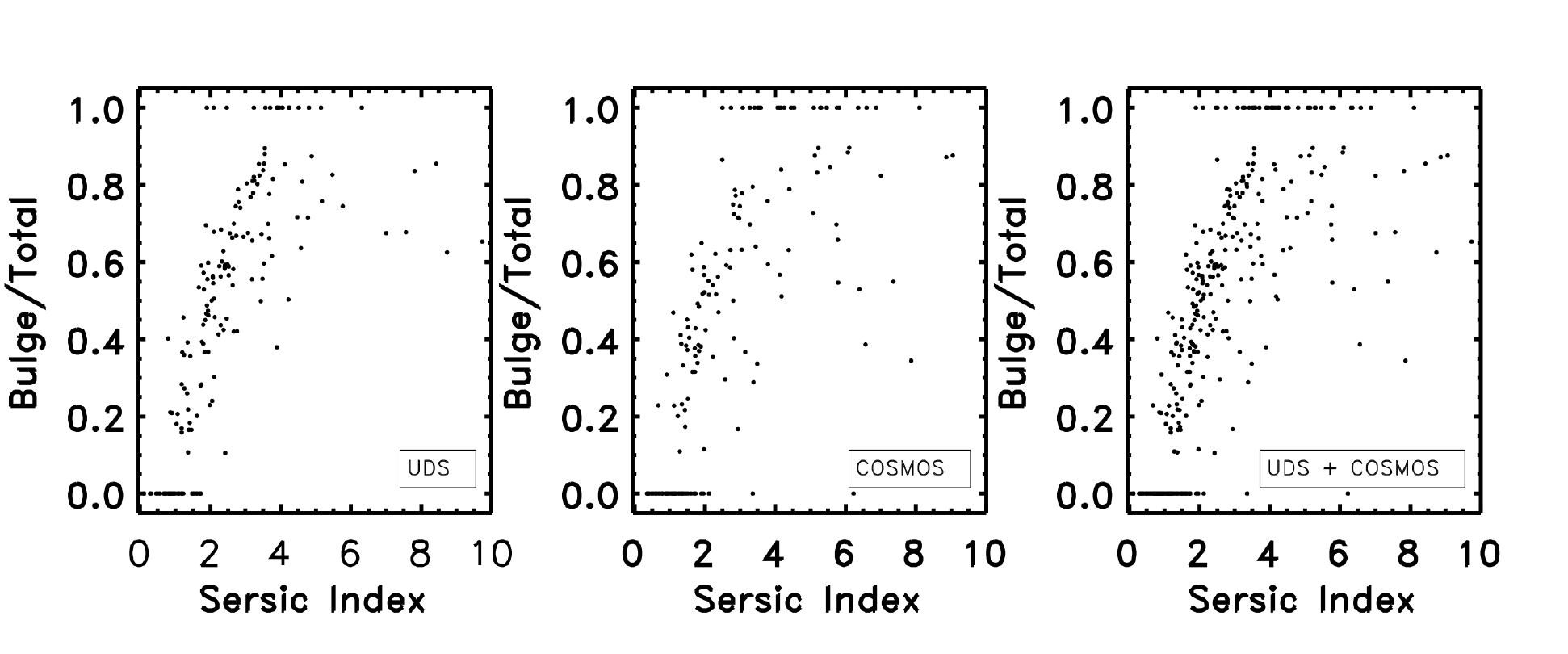}
 \caption[Bulge/Total light fractions vs single-S\'{e}rsic index fits.]{Bulge/Total light fractions against single-S\'{e}rsic index fits, split by field. These plots illustrate that the same trends witnessed in the UDS analysis \citep{Bruce2012} extend to the COSMOS field. }
\label{fig:btsersic}
 \end{center}
\end{figure*}

Having conducted the detailed morphological decomposition described above, and extended this analysis across the four-band wavelength range available from CANDELS, we were then able to compare the overall morphologies fitted by the single-component and the multiple-component fitting techniques. This comparison is shown in Fig.\,\ref{fig:btsersic}, which demonstrates the good correlation between the S\'{e}rsic indices fitted from the single-component models and the bulge/total ($B/T$) light fraction ratios from the $H_{160}$-band multiple-component decompositions. Overall there is a reasonable one-to-one correlation between the $B/T$ light fractions from the multiple-component decompositions and the fitted single-S\'{e}rsic indices (although there is significant scatter), and the S\'{e}rsic index cut at $n=2$ to distinguish between bulge and disk-dominated galaxies closely corresponds to a cut at $B/T=0.5$, with only a few cases where galaxies have $n<2$ and $B/T>0.5$ or $n>4$ but $B/T<0.5$.

\begin{figure*}
\begin{center}
\includegraphics[scale=0.9]{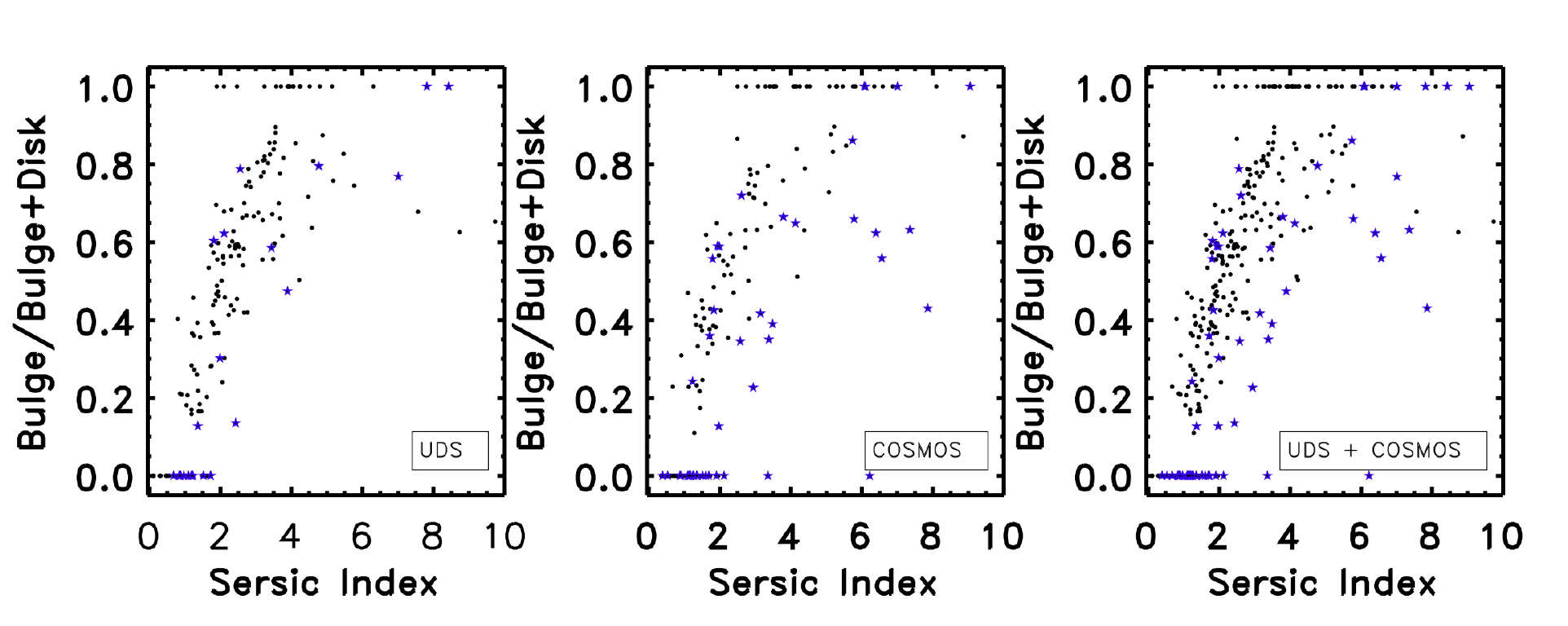} 
 \caption[Bulge/Bulge+Disk light fractions vs single-S\'{e}rsic index fits.]{Bulge/Bulge+Disk light fractions from the $H_{160}$-band modelling against single-S\'{e}rsic index fits, with objects which have best-fit models which contain a PSF component highlighted with blue stars. This demonstrates the same overall trends as in Fig.\,\ref{fig:btsersic}, but here some of the scatter has been reduced and specific cases have been highlighted where the inclusion of a PSF component helps to resolve low $B/B+D$ and high S\'{e}rsic index fits. }
\label{fig:btsersic2}
 \end{center}
\end{figure*}

 In fact, part of this scatter can be explained by the fact that the light fractions are plotted as the bulge/total fraction, where the total light can contain contributions from a PSF.  In this case total=(bulge+disk+PSF), which is not the same as bulge+disk light. For comparison, bulge/bulge+disk fractions are plotted in Fig.\,\ref{fig:btsersic2}, where objects which have a best-fit multiple-component model with a PSF component are highlighted in blue. Adopting this bulge/bulge+disk ratio helps to remove some of the scatter towards high S\'{e}rsic indices but low bulge fractions. In the cases where best-fits contained a PSF, the bulge/total ratios fall as the PSF component has replaced some of the contribution which would otherwise be modelled by the bulge, and the S\'{e}rsic indices are higher because a simple de Vaucouleurs profile no longer provides an adequate fit to these centrally concentrated objects. Thus, for these systems plotting bulge/bulge+disk light ratios arguably provides a characterisation of bulge dominance which is easier to interpret. By highlighting those objects with a significant PSF component ($>10\%$), Fig.\,\ref{fig:btsersic2} also reconciles the single and multiple-component fits for the 2 objects which have $B/T=0$ with $n>2.5$ as it can now be easily seen that these fits have a PSF component. Hence, whilst they have no bulge component these are not ``pure'' disk systems, but have a centrally concentrated light component modelled in the multiple-component analysis by a PSF and in the single-component fits by a high S\'{e}rsic index.
For completeness, we have also examined the correlation between the $B/T$ mass fractions and $n>2.5$. In this case, adopting fractions based on stellar-mass estimates generally increases the contribution from the bulge component, as expected given the different stellar populations comprising the bulge and disk components, but otherwise does not lessen the agreement between the single-S\'{e}rsic index  light-based morphological indicator and the decomposed mass-based discriminator.

The examination of the correlation between single-S\'{e}rsic indices and bulge/total $H_{160}$ light fractions and decomposed stellar-mass estimates, confirms that in the majority of cases the single-S\'{e}rsic index discriminator at $n=2.5$ describes the overall morphologies of these most massive galaxies relatively well, as it provides a good proxy for both light and mass-based measures of $B/T=0.5$. 

\subsection{Size-Mass Relations}

\begin{figure*}
\begin{center}
\includegraphics[scale=0.9]{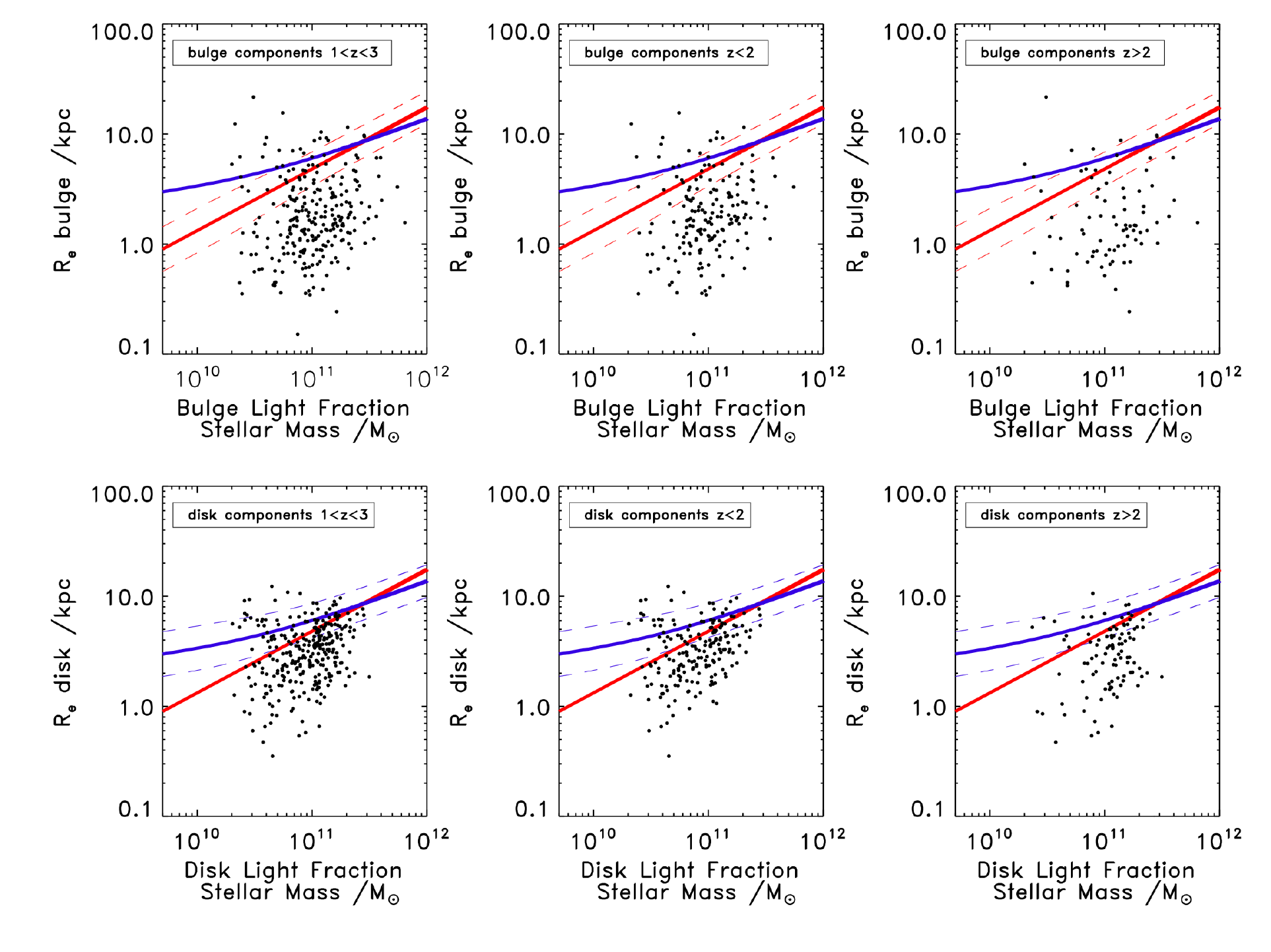} 
 \caption[Size-mass by $H_{160}$ light fractions, Combined UDS + COSMOS.]{Combined UDS + COSMOS size-mass relations for each component, where component masses are determined based on the $H_{160}$-band light fractions. }
\label{fig:sizemassHcom}
 \end{center}
\end{figure*}

The results from the multiple-component decomposition have allowed us to explore how the size-mass relations for the separate bulge and disk components evolve with redshift by accurately decomposing their masses from the multiple-component SED fitting. However, before this is discussed, it is first interesting to explore how the size-mass relations constructed by splitting the mass of each galaxy into each of its separate components according to their contributions to the $H_{160}$-band light compare to the results presented in \citet{Bruce2012}, which used the CANDELS UDS sample alone. The combination of the UDS and COSMOS samples is plotted in Fig.\,\ref{fig:sizemassHcom} and, following the convention in \citet{Bruce2012}, these plots show the size-mass relations for all bulge components in the top panels and disk components in the bottom panels. They are further split by redshift, where the full redshift range ($1<z<3$) is displayed in the far left panels, $1<z<2$ in the middle and $2<z<3$ in the right-hand panels. The bulge relations have been over-plotted with the local \citet{Shen2003} ETG relation in red, with its $1-\sigma$ scatter, and the disk components by the local LTG relation and its scatter, where these relations have been corrected to un-circularised values following the prescriptions outlined in \citet{Bruce2012}. In these plots we only display components with  $M_*>2\times10^{10}\,{\rm M_{\odot}}$, as below this mass the components become sufficiently faint that they may introduce potential biases to the morphological properties fitted, therefore they have been removed from these plots to avoid over-interpretation of sub-components.

It is clear from Fig.\,\ref{fig:sizemassHcom} that the trends reported for the UDS sample are also in place in the COSMOS field, where again we find that the bulge components of massive galaxies display a stronger size evolution with redshift than the disk components. The majority of bulge components have sizes which place them well-below their corresponding local relation, whereas the disk components show a smaller scatter in size with an increased fraction of disks displaying sizes consistent with similarly massive local systems. These results also support the claim of a lower envelope of sizes which scales with mass broadly parallel to the local relation. 

The scatter in the size-mass relation of the bulge and disk components is higher than would be expected from the estimated uncertainties in our size and mass measurements, thus implying a significant intrinsic scatter.

\begin{figure*}
\begin{center}
\includegraphics[scale=0.9]{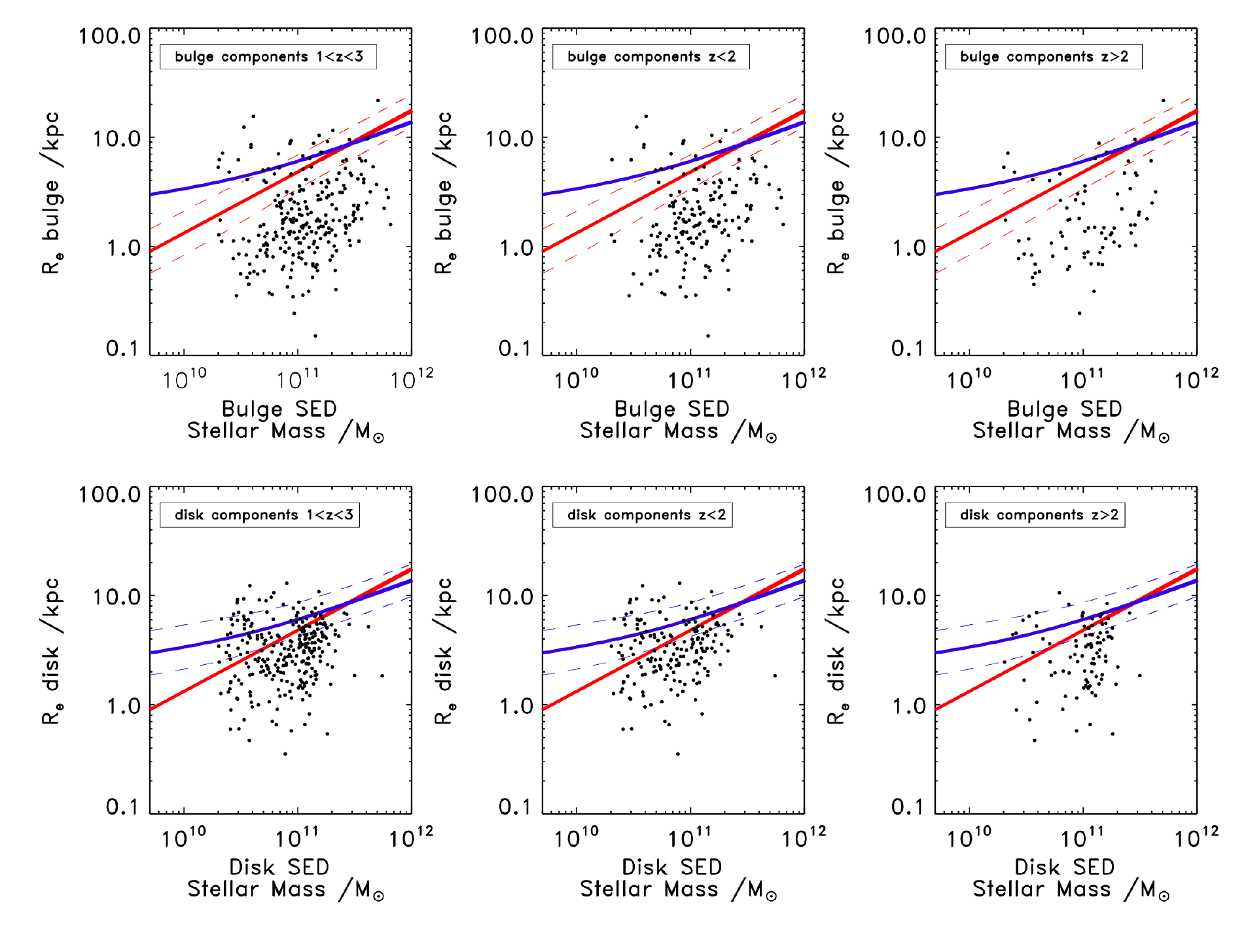} 
\end{center}
 \caption[Size-mass by SED component mass, combined]{Combined UDS + COSMOS size-mass relations for each component, where now component masses are estimated from the multiple-component SED fitting. The same trends in the sizes of the components witnessed for the relations constructed using masses split according to the $H_{160}$-band light fractions are also displayed by these relations, which adopt the more robust decomposed SED-fitted component masses. This includes the larger fraction of bulge components which lie below the local relation and the smaller median sizes compared to the disk components, in addition to the lower envelope of sizes displayed. }
\label{fig:sizemasssedcom}
\end{figure*}

We now move on to consider the size-mass relations based on the separate component masses estimated from SED fitting. These results are presented in Fig.\,\ref{fig:sizemasssedcom}. Comparison between these size-mass relations and those plotted in e.g. Fig.\,\ref{fig:sizemassHcom} reveals no significant change in the reported relations for either the bulge or disk components. This suggests that the simplified approach of attributing masses to each component based on their contributions to the $H_{160}$-band light fractions provides, at least on average, a good proxy for SED fitted stellar-mass decompositions.

This stellar mass decomposition confirms all of the morphological trends revealed by the previous light-fraction decomposed size-mass relations, including the stronger evolution witnessed for bulge components over disks, both in terms of the number of bulges which fall below their respective local relations, and in the median sizes of the populations in both the $z<2$ and $z>2$ redshift bins. These results are summarised in Table \ref{table:fits1}, which shows that, within the errors, these trends are consistent across both fields, and are in agreement with the statistics quoted in \citet{Bruce2012} for the size-mass relations from masses based on $H_{160}$-band light fractions for the UDS field alone. Again, the uncertainties on these values do not allow us to draw any robust conclusions about the change in these fractions with redshift, although we do note that the CANDELS-COSMOS sample contains a larger number of bulge-dominated objects at $z>2$ which is responsible for the rise of bulges with sizes comparable to local ETGs within this redshift bin. This could be produced by a systematic error (focussing) in the determination of the photometric redshifts. However, the effects of redshift focussing have been studied in both fields with the (albeit low numbers of) spectroscopic redshifts available, but we find no strong evidence for this effect amongst the bulges and conclude that this larger abundance of high-redshift bulges in the CANDELS-COSMOS field may be due to an interception of genuine structure in the COSMOS field (although no obvious spatial clustering of these objects is seen).

\begin{table*}
\begin{center}
\begin{tabular}{m{2cm}m{3.0cm}m{2cm}m{2cm}m{2cm}}
\hline
&
&
$1<z<3$ &
$1<z<2$&
$2<z<3$ \\
\hline
COSMOS&
bulges on &
$21\pm4\%$ &
$14\pm4\%$ &
$36\pm8\%$ \\
&
bulges below&
$79\pm4\%$ &
$86\pm4\%$ &
$64\pm8\%$ \\
&
disks on &
$35\pm4\%$ &
$39\pm5\%$ &
$25\pm8\%$ \\
&
disks below &
$65\pm4\%$ &
$61\pm5\%$ &
$75\pm8\%$ \\
\hline
UDS&
bulges on &
$15\pm3\%$ &
$16\pm4\%$ &
$12\pm5\%$ \\
&
bulges below &
$85\pm3\%$ &
$84\pm4\%$ &
$88\pm5\%$ \\
&
disks on &
$56\pm4\%$ &
$59\pm5\%$ &
$52\pm6\%$ \\
&
disks below &
$44\pm4\%$ &
$41\pm5\%$ &
$48\pm6\%$ \\
\hline
Combined&
bulges on &
$18\pm2\%$ &
$15\pm3\%$ &
$23\pm5\%$ \\
&
bulges below &
$82\pm2\%$ &
$85\pm3\%$ &
$77\pm5\%$ \\
&
disks on &
$47\pm3\%$ &
$49\pm4\%$ &
$43\pm5\%$ \\
&
disks below &
$53\pm3\%$ &
$51\pm4\%$ &
$57\pm5\%$ \\
\hline
\end{tabular}
\caption{The fractions of components which lie on (or above) their respective local relations within the $1-\sigma$ scatter and below the $1-\sigma$ scatter of their relations, where masses for each component have been estimated separately from the multiple-component SED fitting.}
 \label{table:fits1}
\end{center}
\end{table*}

From this discussion it is evident that the adoption of the more rigorous SED decomposed component masses, over the $H_{160}$-band light fraction mass decompositions, has not significantly influenced the positions of components in their respective size-mass relations, nor altered the basic trends reported. However, the full SED stellar-mass decomposition not only provides robust individual component masses, but also delivers estimates of the star-formation activity of each object.

\subsection{Star-formation trends}

Early size-mass studies (e.g. \citealt{Kriek2006,Toft2007}) reported a correlation between compactness and passivity which has since gained substantial support in the literature, but these studies are not only limited to morphological classifications based on single-S\'{e}rsic index fits but also (with the exception of IFU spectroscopic studies) global star-formation rates. 
In this respect, the advantage of our full SED multi-band decomposition technique becomes clear, as it has allowed us to estimate the SFR for each individual bulge and disk component.

\begin{figure*}
\begin{center}
\includegraphics[scale=0.9]{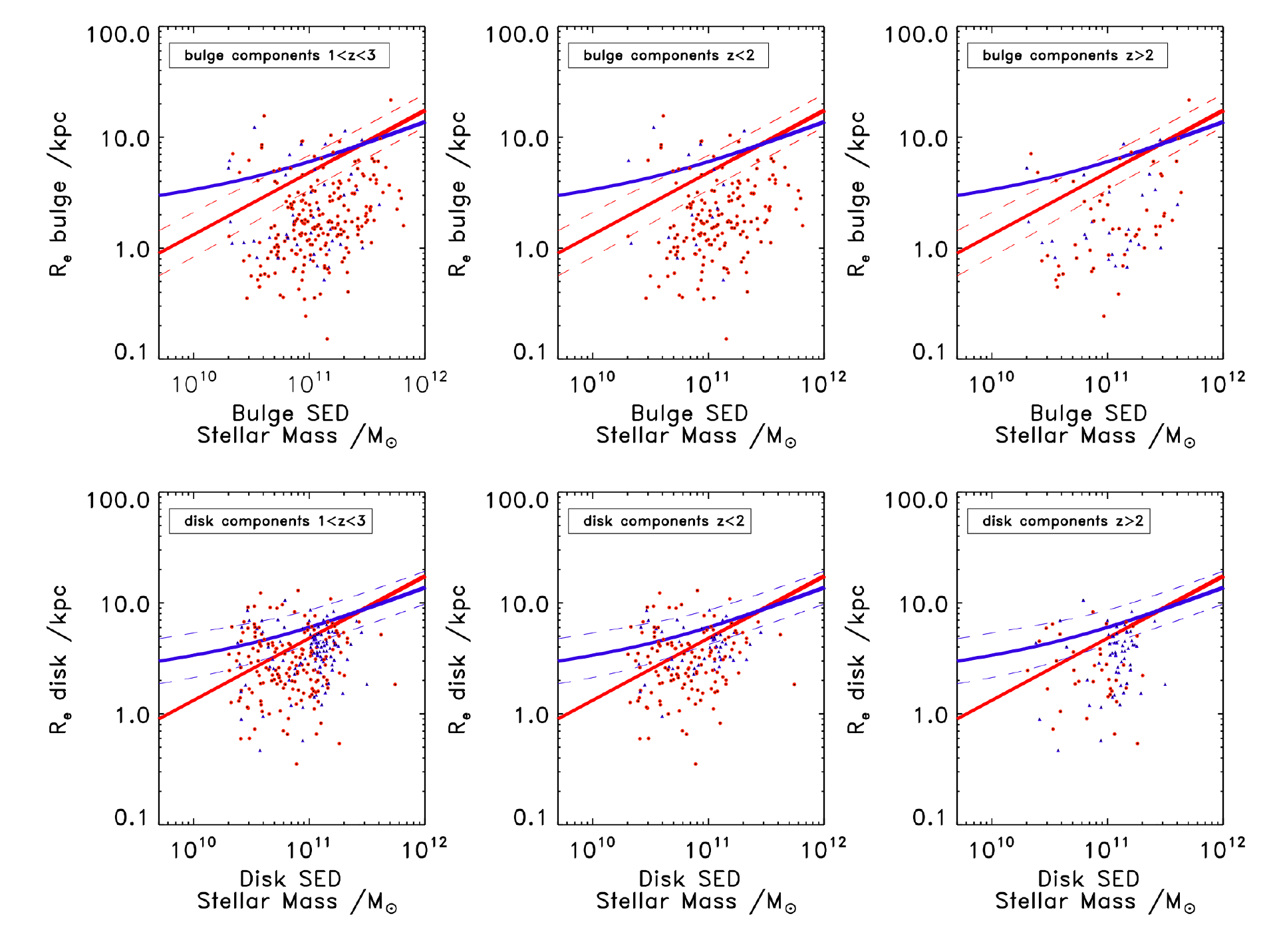}
 \caption[Size-mass by SED mass and component sSFRs, combined.]{Combined UDS + COSMOS size-mass relations for each component, where component masses are estimated from the multiple-component SED fitting and components are coloured by their star-formation activity using the $sSFR<10^{-10}\,{\rm yr^{-1}}$ limit for passivity (red circle are passive components and blue triangles are star-forming components), and based on the separate component sSFRs from the multiple-component SED fitting. These coloured relations do not reveal a clear division in the sizes of passive and star-forming components, but instead show that the star-forming and passive bulges have comparable sizes with some of the largest bulges being passive and some of the most compact bulges displaying evidence of on-going star formation.}
\label{fig:sfrcom}
\end{center}
\end{figure*}

In Fig.\,\ref{fig:sfrcom}, we now show the size-mass relations plotted with the SED-fitted decomposed stellar masses are now coloured by their individual component star-formation rates, where for simplicity we have adopted the $sSFR=10^{-10}\,{\rm yr^{-1}}$ discrimination between star-forming and passive components and plot the passive components in red and the star-forming components in blue. These size-mass plots, containing separate component star-formation information, are shown in  
Fig.\,\ref{fig:sfrcom}. These plots do not immediately display a clear division between the sizes of passive and star-forming components, but instead reveal that a fraction of the most compact bulges and disks display signs of continued star-formation, while some of the largest bulges and disks are classified as passive. The evolution of bulge and disk components split into their star-forming and passive populations is studied in more detail in the following section by exploring how the median sizes of each sub-population, given as a fractional size of their local counter-parts, evolve with redshift.

\begin{table*}
\begin{center}
\begin{tabular}{m{4.0cm}m{2cm}m{2cm}m{2cm}}
\hline
&
$1<z<3$ &
$1<z<2$&
$2<z<3$ \\
\hline
sf bulges on &
$29\pm7\%$ &
$40\pm13\%$ &
$23\pm8\%$ \\
sf bulges below&
$71\pm7\%$ &
$60\pm13\%$ &
$77\pm8\%$ \\
sf disks on &
$48\pm5\%$ &
$54\pm7\%$ &
$42\pm7\%$ \\
sf disks below &
$52\pm5\%$ &
$46\pm7\%$ &
$58\pm7\%$ \\
passive bulges on &
$15\pm2\%$ &
$13\pm3\%$ &
$23\pm6\%$ \\
passive bulges below&
$85\pm2\%$ &
$87\pm3\%$ &
$77\pm6\%$ \\
passive disks on &
$46\pm4\%$ &
$46\pm4\%$ &
$45\pm8\%$ \\
passive disks below &
$54\pm4\%$ &
$54\pm4\%$ &
$55\pm8\%$ \\

\hline
\end{tabular}
\caption{The fractions of the combined UDS and COSMOS sample components which lie on or below their respective local relations, where masses for each component have been estimated separately from the multiple-component SED fitting, split further into their star-forming and passive populations using the individual component sSFRs.}
\label{table:fits2}
\end{center}
\end{table*}

\subsection{Fractional Size Evolution}

In order to better explore the evolution of bulge and disk components split into their star-forming and passive populations, the fractions of each population which display sizes consistent with or below their respective local relations are given in Table \ref{table:fits2}, along with the offsets of the median sizes of these populations from their local relations in Table \ref{table:relations}. For clarity all bulges have been compared with the local ETG relation, and disks with the local LTG relation \citep{Shen2003}.

\begin{figure}
\begin{center}
\includegraphics[scale=0.54]{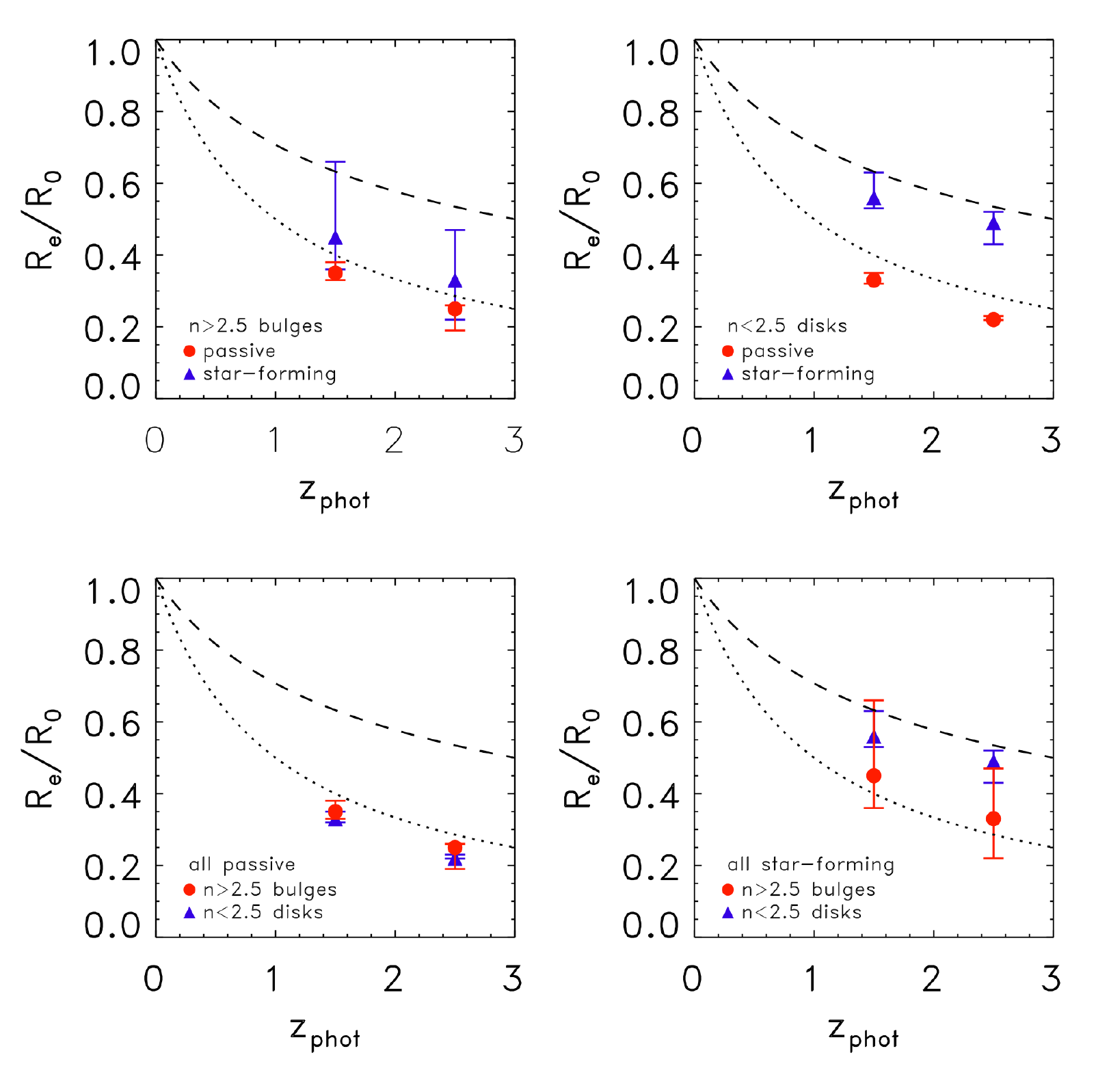}
 \caption[Fractional ETG and LTG size evolution]{The fractional size evolution of galaxies classified as ETGs and LTGs based on a cut at $n=2.5$ for our single-S\'{e}rsic index fits. The top panels are split into all ETGs (left) and all LTGs (right), whereas the bottom panels show all passive galaxies (left) and all star-forming galaxies (right) to allow an easier comparison of the same data depending on the morphological or star-formation activity distinctions. Over-plotted as the dotted line is the fitted $R_{e}(z)/R_{0}\propto(1+z)^{-1}$ ETG size evolution from \citet{Vanderwel2008}, and the dashed line is the relation fitted to our decomposed star-forming disk sample (top-right panel of Fig.\,\ref{fig:offsets1}) given by $R_{e}(z)/R_{0}\propto(1+z)^{-0.5}$. Using the single-S\'{e}rsic fits, the passive disks are as compact as star-forming and passive bulges, and are significantly smaller than the sizes of star-forming disks. The sizes of the passive and star-forming bulges are equally compact within the errors, and despite the larger uncertainties, this trend remains for the multiple-component SED decompositions represented in Fig.\,\ref{fig:offsets1}.}
\label{fig:offsets2}
\end{center}
\end{figure}

\begin{figure}
\begin{center}
\includegraphics[scale=0.54]{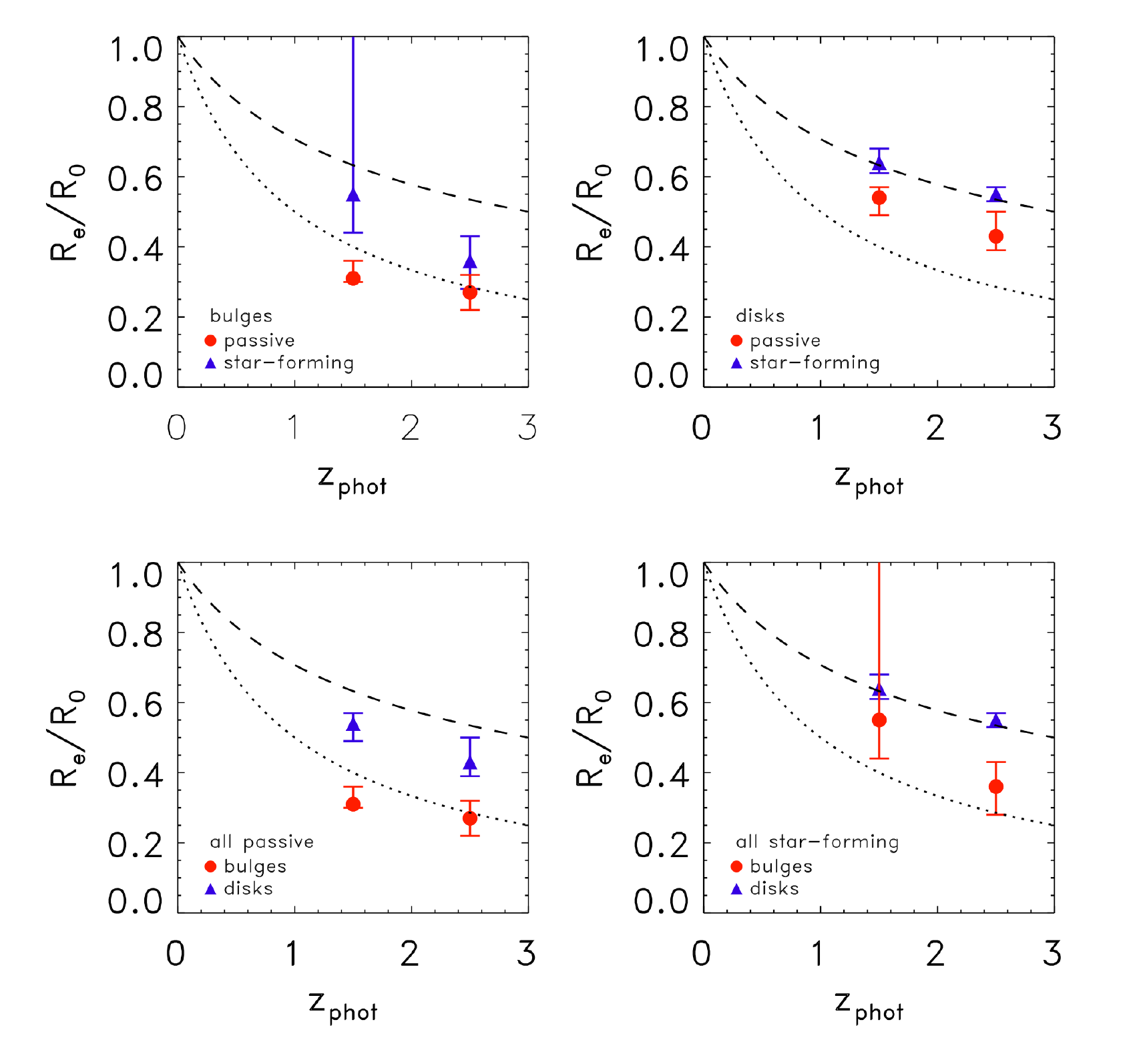}
 \caption[Fractional bulge and disk size evolution]{The fractional size evolution of all the bulge and disk components with respect to their local relations in the top-left and right panels, respectively, and for all passive and star-forming components in the bottom-left and right panels. In this case star-forming and passive disks have been compared to the local LTG relation and star-forming and passive bulges have been compared to the local ETG relation. Over-plotted as the dotted line is the fitted $R_{e}(z)/R_{0}\propto(1+z)^{-1}$ ETG size evolution from \citet{Vanderwel2008}, and the dashed line is the relation fitted to my decomposed star-forming disk sample (top-right panel) given by $R_{e}(z)/R_{0}\propto(1+z)^{-0.5}$. Using the multiple-component SED decompositions, the passive and star-forming bulges arguably remain equally compact within the large errors, but passive disks display an intermediate size as they are larger than their bulge counterparts but smaller than the star-forming disks. }
\label{fig:offsets1}
\end{center}
\end{figure}

These results reveal that the sizes of passive and star-forming bulges are consistently compact, within the errors, and that star-forming disks are significantly larger. However, they also show that passive disks have intermediate sizes, larger than their passive bulge counterparts, but smaller than the disks which remain active. To better explore these results we have calculated the sizes of these sub-divided populations as a fraction of the present-day sizes of similarly massive galaxies, using the median fractional sizes of all objects. 
This assumes that the slope of the size-mass relation is constant over $0<z<3$ (e.g. \citealt{McLure2013,vanderWel2014}). These fractional sizes for the bulge and disk components from the full SED-fitting decomposition are shown in  Fig.\,\ref{fig:offsets1}. This confirms the trends determined from the size-mass relation plots, but also allows for a more direct and intuitive comparison of component sizes split into star-forming and passive populations at different redshifts. In Fig.\,\ref{fig:offsets2} we have also included the fractional size evolution as determined from the single-S\'{e}rsic fitting, as all previous $1<z<3$ light-profile fitting size-mass studies have relied on this parameter to distinguish between bulge and disk-dominated systems. Thus, it allows not only a direct comparison with previous literature results, but also with the multiple-component SED-fitting decomposition results and so serves to highlight the additional insight which can be gained from adopting the decomposition method for galaxy size measurements.

\begin{table}
\begin{center}
\begin{tabular}{m{1.7cm}m{1.9cm}m{1.9cm}m{1.7cm}}
\hline
&
$1<z<$3&
$1<z<$2&
$2<z<$3\\
\hline
bulge
\newline components
\newline &
$3.09\pm0.20$ &
$2.93\pm0.32$&
$3.41\pm0.58$\\
star-forming
\newline bulges 
\newline&
$2.81\pm0.64$&
$1.83\pm0.30$&
$3.81\pm1.0$\\
passive
\newline bulges
\newline&
$3.01\pm0.19$&
$3.00\pm0.14$&
$3.24\pm0.44$\\
\hline
disk 
\newline components
\newline &
$1.77\pm0.10$&
$1.65\pm0.14$&
$1.99\pm0.25$\\
star-forming
\newline disks
\newline &
$1.62\pm0.15$&
$1.50\pm0.13$&
$1.78\pm0.20$\\
passive
\newline disks
\newline &
$1.94\pm0.25$&
$1.72\pm0.27$&
$2.35\pm0.41$
\end{tabular}
\caption[Offsets from local relations]{The fractional offsets of the median sizes of each population from their respective local relations.}
\label{table:relations}
\end{center}
\end{table}

Starting with  Fig.\,\ref{fig:offsets2} for the single-S\'{e}rsic fitting technique, where disk-dominated galaxies are classified as $n<2.5$ following the convention of \citet{Shen2003} and bulges as $n>2.5$ , we found that the size of passive bulges, passive disks and active bulges are all consistent within their errors and are similarly compact, but that star-forming disks are significantly larger. This can clearly be seen in Fig.\,\ref{fig:offsets2}, where we have over-plotted as the dotted line the size evolution for ETGs as fitted by \citet{Vanderwel2008}, given by $R_{e}(z)/R_{0}\propto(1+z)^{-1}$, and as the dashed line the fitted size-evolution of the decomposed star-forming disks (top-right panel of Fig.\,\ref{fig:offsets1}) , as given by $R_{e}(z)/R_{0}\propto(1+z)^{-0.5}$. 

These trends are consistent with previous studies such as \citet{McLure2013}, but raise questions over the mechanisms by which star-forming galaxies quench and also significantly reduce in size to form the passive-disk population. One possible reason for this apparent discrepancy may be that the passive-disk galaxies are more bulge-dominated than the star-forming disks, and are therefore biased to smaller sizes in this comparison.

 In order to test this we have explored the S\'{e}rsic index distributions of both the passive and star-forming disks and do find that using a cut at $n=2.5$, the passive disks are centred on a higher n values than the star-forming disks. As a result, we have experimented with decreasing the S\'{e}rsic index value used as the discriminator between bulges and disks, in an attempt to ensure that in order to be classified as passive disks these galaxies are as disk-dominated as possible. By decreasing the S\'{e}rsic index cut to $n=2$ and $n=1.5$ we find a better agreement between the S\'{e}rsic index distributions for the passive and star-forming disks (although the passive disks are still centred on slightly higher values of $n$), but this does not affect the derived fractional sizes of this population. Thus, from the single-S\'{e}rsic fitting technique one would always find that the star-forming disks are substantially larger than the passive disks, and in fact that the passive disks have sizes comparable to the star-forming and passive bulges.

Adopting the multiple-component SED-fitting decompositions yields the fractional size evolutions displayed in Fig.\,\ref{fig:offsets1}, which have again been over-plotted with the \citet{Vanderwel2008} ETG ($R_{e}(z)/R_{0}\propto(1+z)^{-1}$) and the fitted star-forming disk  ($R_{e}(z)/R_{0}\propto(1+z)^{-0.5}$) relations.
This shows similar size evolution for the bulge components and star-forming disks, but now reveals that the passive disks now have an intermediate size, between the passive and star-forming bulges and the star-forming disks. 

It is possible that the inclusion of all passive disk components in this sample introduces some effects associated with the lower masses that are being probed, as for the single-S\'{e}rsic index fits all bulges or disks have stellar masses $M_*>10^{11}\,{\rm M_{\odot}}$, but the decomposed component masses can range as low as $M_*=2\times10^{10}\,{\rm M_{\odot}}$. For these low-mass components we then compared their sizes to similarly massive local galaxies via the \citet{Shen2003} LTG relation, but in this case we are comparing the size of a low-mass disk component of a much more massive bulge-dominated galaxy to a low-mass disk-dominated system at low redshift, which may bias the fractional size measurements of these galaxies to higher values. However, to account for this we examined these relations using only the bulge component of bulge-dominated galaxies and the disk component of disk-dominated galaxies. Whilst the adoption of this subset does not significantly affect the fractional sizes of the bulges or star-forming disks, it does reduce the size of the passive disks, although not by an amount which makes them consistent with the $n<2.5$ single-S\'{e}rsic index passive galaxies. Hence, even though there may be some effect from low-mass sub-components which drives the passive disks to larger sizes, it is not the dominant reason for the increase in passive disk sizes from the multiple-component SED-fitting decomposition, and we are left to conclude that the more accurate decomposition of both individual component stellar masses and star-formation rates reveals a potential bias in the results from the single-S\'{e}rsic index fitting technique, with passive disk components genuinely having an intermediate size.

\section{Discussion}
\subsection{Comparison to results in the literature}

Despite the profusion of morphological studies of massive $z>1$ galaxies over the past 5-10 years, these studies have adopted a number of different selection criteria and are often biased towards selecting passive or early-type systems. As a result, it has been difficult to disentangle the trends for galaxies with early-type (ETG) morphologies to be more compact than late-type (LTG) systems, and for passive galaxies to be more compact than those which display on-going star-formation. In order to conduct a robust and direct comparison between the size evolution of bulge and disk-dominated galaxies, sub-divided further into their passive and star-forming populations, it is important to adopt an unbiased, mass-selected sample, as has been done for this work. 
Here we limit comparison of our results to several of the most notable studies which have well-defined samples and stellar-mass and size determination procedures most directly comparable with our own.

We first draw a comparison between our single-S\'{e}rsic fits and the study of \citet{Buitrago2008} conducted at $2<z<3$ for 82 $M_*>10^{11}\,{\rm M_{\odot}}$ galaxies  with NICMOS {\it HST} imaging. In this study, \citeauthor{Buitrago2008} split their sample into bulge and disk-dominated systems using a S\'{e}rsic index cut at $n=2$ and find that, on average: the $n<2$ disks have a fractional size $R_{e}/R_{0}=0.38\pm0.05$; and bulge systems with $n>2$ have $R_{e}/R_{0}=0.23\pm0.04$. Within the errors, these results are consistent with our $z>2$ sample using the similarly modelled single-S\'{e}rsic fits, where as discussed previously, cutting our sample at the lower $n=2$ limit does not affect the median fractional sizes that we determine. 

We can also compare with the study of \citet{McLure2013} for $M_*>6\times10^{10}\,{\rm M_{\odot}}$, $z=1.4$ galaxies in the UDS covered in the K-band by UKIDSS UDS, and with spectra from FORS2. \citet{McLure2013} split their mass-selected sample by both morphology, above and below $n=2.5$, and in terms of the overall galaxy star-formation activity. They report that, at this redshift, $n<2.5$ disks have a median $R_{e}/R_{0}=0.465\pm0.032$ and  $n>2.5$ bulges have $R_{e}/R_{0}=0.42\pm0.05$. Whereas, splitting by star-formation activity, their passive galaxies have a median $R_{e}/R_{0}=0.42\pm0.035$ and the star-forming sample have a median $R_{e}/R_{0}=0.625\pm0.078$. 

\citet{McLure2013} comment that the apparent difference in size between their star-forming and $n<2.5$ disk samples may be due to the contribution of a significant fraction of passive disks to the median size offsets. In comparison, both our $z<2$ single-S\'{e}rsic and multiple-component fits are roughly consistent with the results from this study, although we note that the fractional sizes of the \citet{McLure2013} star-forming and passive samples are more consistent with our multiple-component fits than our single-S\'{e}rsic results. This may in part be due to the fact that although \citeauthor{McLure2013} adopt the same $sSFR<10^{-10}{\rm yr^{-1}}$ passivity criterion, their adoption of ``double-burst'' star-formation histories during SED fitting, may account for the better agreement between their passive and star-forming fractional size measurements and our decomposed fits (which have also allowed multiple star-formation history components for each galaxy).

We next compare our results to the study of \citet{Toft2007}, which was among the first to note the correlation between galaxy passivity and compactness. The \citet{Toft2007} study was conducted at $z\approx2.5$ using {\it HST} NICMOS and ACS imaging, and classified galaxies as active or passive depending on whether or not the SED fits to the galaxies were better modelled by constant or burst star-formation histories, and were then cross-checked with $24\mu$m data. \citet{Toft2007} report that : at $z=2.5$ passive galaxies have  $R_{e}/R_{0}=0.19\pm0.03$ and star-forming galaxies have $R_{e}/R_{0}=0.45\pm0.15$.

 Again, these results are broadly consistent with the passive and star-forming fractional size estimates from both our single-S\'{e}rsic and decomposed fits within the errors, especially given the different classifications adopted for star-forming and passive galaxies and that the \citet{Toft2007} sample spans a much wider, and lower mass range ($0.4\times10^{10}<M_*<5.5\times11^{11}\,{\rm M_{\odot}}$). It should also be noted that the \citet{Toft2007} passive sample has a S\'{e}rsic index distribution centred on $n<4$, with $\approx 80\%$ of objects being better fit with $n=1$ rather than $n=4$ light profiles.

Finally, in order to complete the literature comparison we consider the study of \citet{Cimatti2008} at $1.4<z<2$ for a spectroscopically confirmed passive GMASS sample imaged with {\it HST} NICMOS and ACS.\citet{Cimatti2008} split their sample into two redshift bins and report that at $z=1.6$ their passive galaxies have a median $R_{e}/R_{0}=0.37\pm0.08$ and at $z=2.5$ $R_{e}/R_{0}=0.29\pm0.14$. Again, these results are in general agreement with our single-S\'{e}rsic fits, but in this case, as to some extent with the study of \citet{Toft2007}, a departure between the size of passive disks and those of the passive bulges begins to become more apparent.

From this comparison with previous studies which have split their sample according to (or various combinations of):  i) $n=2$; ii) $n=2.5$; iii) photometrically or spectroscopically determined star-formation rates, there is clear evidence for the trends for: i) passive galaxies at any redshift to be more compact than star-forming galaxies,  ii) ETGs ($n>2.5$) at any redshift to be more compact than LTGs ($n<2.5$); iii) star-forming galaxies to display a shallower size evolution with redshift than passive galaxies.
However, the intermediate sizes of passive disks only become fully apparent from the morphological decompositions presented in this work. 

As a consequence, this suggests that compactness may correlate with some combination of passivity {\it and} the presence of a significant bulge component in the galaxy.

\subsection{Insights into galaxy evolution}

\begin{figure}
\begin{center}
\includegraphics[scale=0.54]{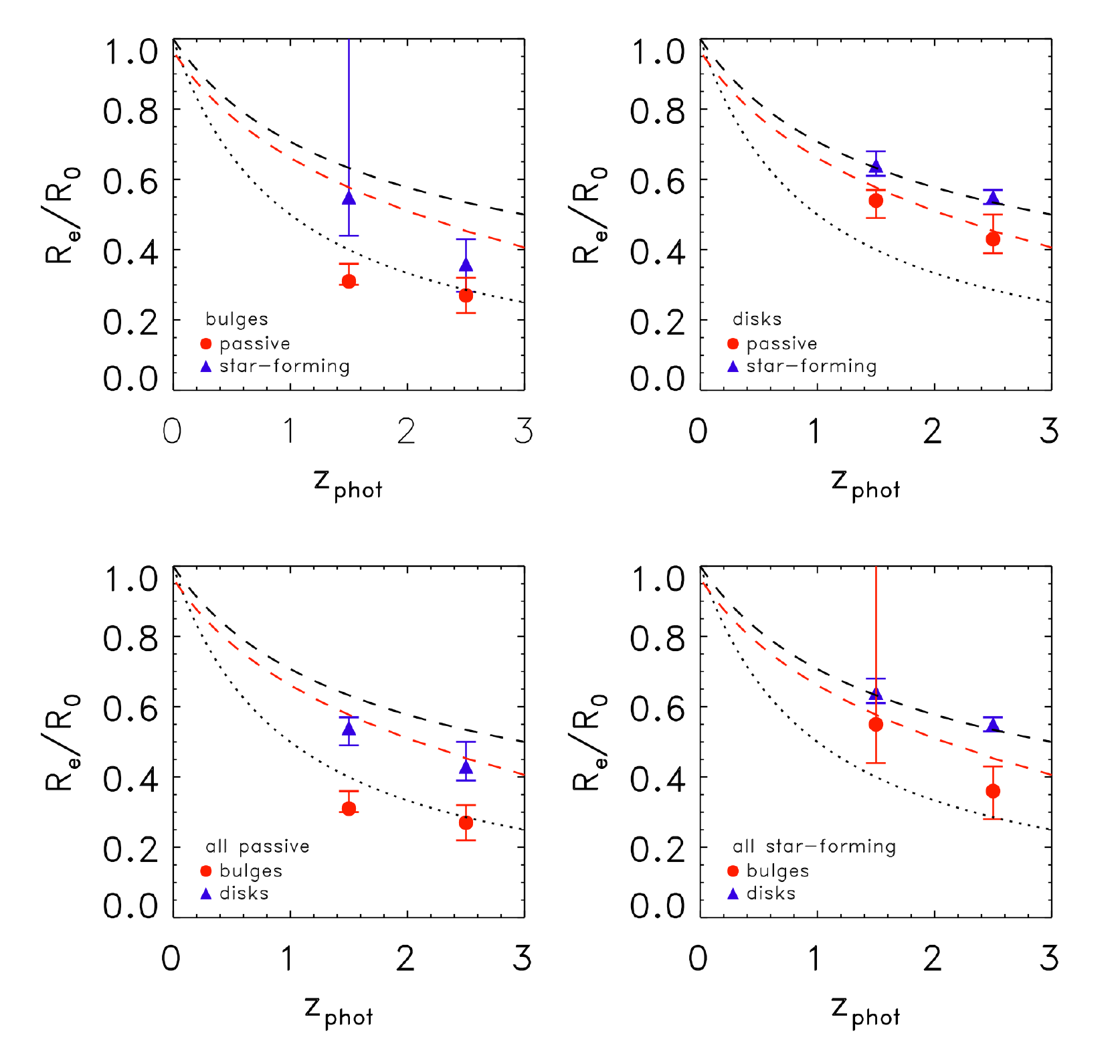}
 \caption[Fractional bulge and disk component size evolution including the relation for progenitors of the passive disks ]{Fractional bulge and disk component size evolution now over-plotted in the dashed red line by the relation for the progenitors of the passive disks to allow direct comparison between the sizes of the passive disks and their 1Gyr earlier star-forming progenitors.}
\label{fig:offsets7}
\end{center}
\end{figure}

By extending our multiple-component light-profile fitting to multi-band photometry and SED fitting to provide individual component masses and star-formation rates, we have directly shown that the median sizes of passive disks are smaller than those of star-forming disks, which raises questions of how these star-forming disks evolve into the passive population.
In order to better understand this evolution, it is important to note that it is not necessarily physically meaningful to compare the sizes of passive and star-forming disks at the same redshifts. In a secular evolution scenario we expect the star-forming disks to evolve into the passive population, therefore it is more meaningful to compare the star-forming disks at higher redshifts to passive disks at lower redshifts.
In order to conduct this comparison we have used the SED fits of the passive disks to reverse engineer their fitted star-formation histories back to the point at which they would last be classified as star-forming, given our $sSFR>10^{-10}{\rm yr^{-1}}$ criterion, and determine the time that the component has been quenched as the difference between the age of the galaxy at the best-fit redshift and the time when it was last active. We find that the majority of components were last active $\simeq1$ Gyr before the epoch of observation. Therefore, in order to best compare between the sizes of passive disks and their star-forming progenitors, the comparison should be conducted between passive disks at their current redshift and star-forming disks at redshifts which correspond to $\sim1$ Gyr earlier. This is shown in Fig.\,\ref{fig:offsets7}, where the fitted $r_{e}\propto(1+z)^{-0.5}$ star-forming disk relation has been re-plotted for redshifts corresponding to $\sim1$ Gyr earlier and can be directly compared to the sizes of the passive disks. From this plot it can be concluded that the sizes of the passive disks at $1<z<3$ are consistent with their star-forming progenitors. It is also worth noting that the most recently quenched disks, (last active $<0.5$\,Gyr earlier), have a median size of $2.47^{+0.28}_{-0.2}$\,kpc, which is larger than the median of $1.94^{+0.09}_{-0.14}$\,kpc for the whole passive disk population. This lends further support to the assertion that the size offset between the passive and star-forming disks can be accounted for by the relation between size and the redshift of quenching.

 \begin{figure}
\begin{center}
\includegraphics[scale=0.58]{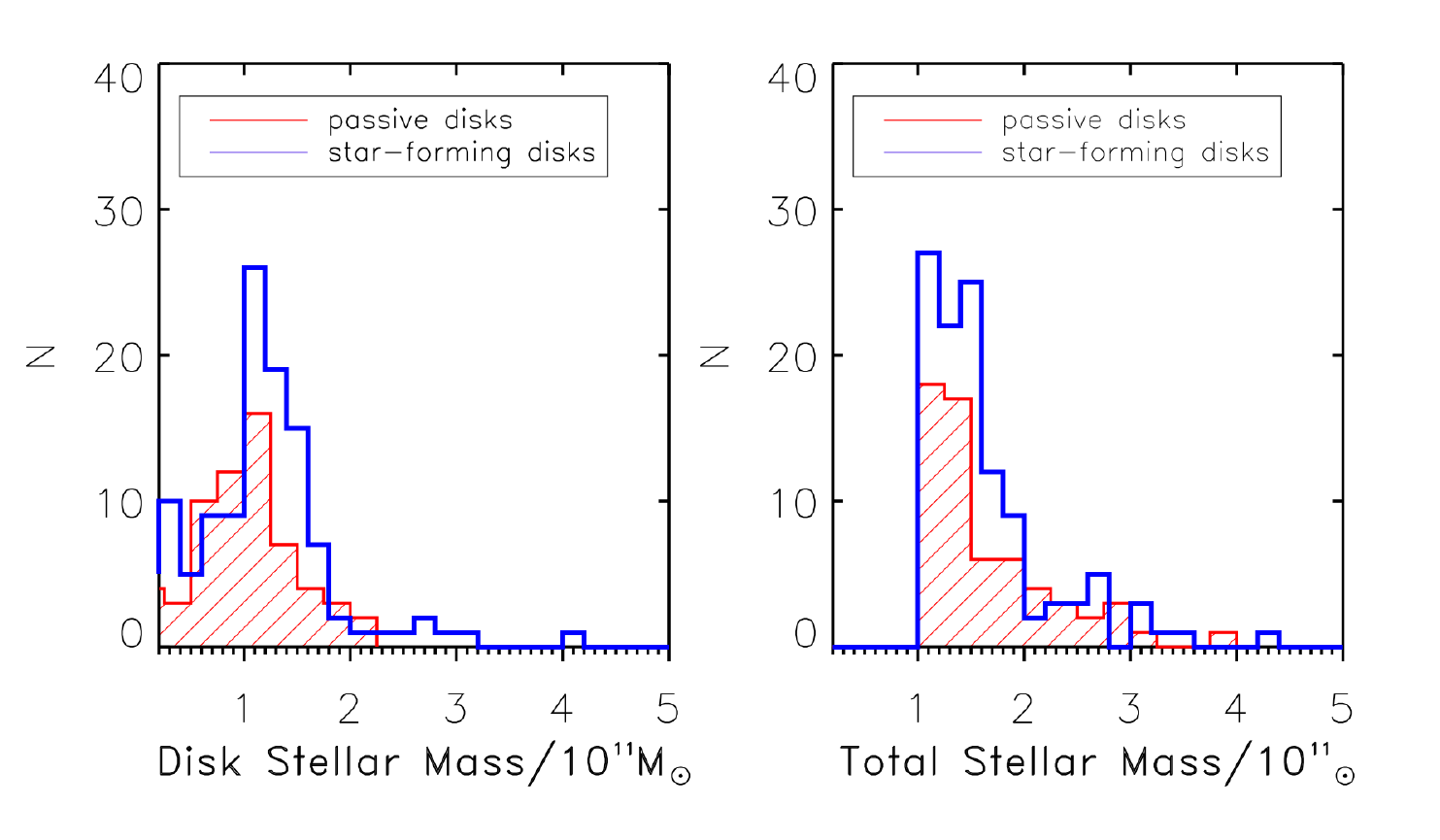}
 \caption[Distributions of the masses of the passive and star-forming disks. ]{The distribution of the disk component (left) and total (right) masses for the passive and star-forming disk-dominated galaxies. While the disk component masses have a probability $p=0.06$ of being drawn from the same distribution, with the star-forming progenitors appearing to have a distribution centred on higher masses, the total galaxy masses for these passive and star-forming disk-dominated systems are more comparable with $p=0.66$. This is consistent with the secular quenching scenario as the evolution of these systems may be accompanied by a transfer of mass from the disk to the bulge components, which would reduce the mass in the disk components but leave the total galaxy mass unchanged.}
\label{fig:offsets9}
\end{center}
\end{figure}

The distributions of the masses in the disk components of these disk-dominated passive galaxies and their star-forming progenitors are shown in Fig.\,\ref{fig:offsets9}, alongside the distributions of the total galaxy masses. The disk component masses of the passive and star-forming disks have a rather low probability of being drawn from the same distribution ($p=0.06$ from a K-S test), with the star-forming disks appearing to have a distribution centred on slightly higher stellar masses, while the total galaxy mass distributions are more comparable ($p=0.66$). Any potential evidence for the star-forming disk components being more massive than the passive disk components which they evolve into is in fact  consistent with the secular quenching scenario as the evolution of these systems may be accompanied by a transfer of mass from the disk to the bulge components (e.g. \citealt{Bournaud2011}), which would reduce the mass in the disk components but leave the total galaxy mass unchanged. 
Whether the processes which quench star-formation are secular or merger driven, these observations challenge models to account for this mass evolution and both the presence of massive, quenched disks and their sizes.

When considering the size evolution of all the individual components it is interesting to address the current claims in the literature that the size evolution of passive galaxies from $z\approx3$ to the present day can be better explained by the addition of newly-quenched, larger galaxies to this population with time (where the size of newly-quenched, younger, galaxies scales with the average density of the Universe at the epoch when they quenched) (e.g. \citealt{Valentinuzzi2010b,Cassata2011,Poggianti2013, Poggianti2013b,Cassata2013,Carollo2013, Krogager2013}), than by the evolution in size of individual galaxies. One of the natural predictions of this scenario is the star-formation dependent size of \emph{both} bulge and disk components, as at any given epoch the star-forming components are expected to be larger, given the fact that they have not yet quenched but do so at later times. While there is evidence for this trend in the disk components, we do not find strong evidence for a size offset between the passive and active bulges, as has been previously reported by, for example, \citet{Carollo2013}, albeit for lower-mass systems. However, as discussed in Bruce et al. (2014a), the star-forming bulge population is subject to significant contamination from sub-dominant active disks, and the scatter in the sizes of these components is large.

Finally, we discuss the sizes of our passive disks and star-forming bulges within the context of the evolution scenario suggested by \citet{Barro2013,Barro2013b} and \citet{Dekel2014}. These studies propose that extended star-forming disks at high redshifts first undergo gas rich dissipational major mergers or experience violent disk instabilities, which both shrink their sizes and transform them from disk to bulge systems, and are then subsequently quenched (with passive bulges later growing in size via e.g. minor mergers). 
The compact sizes of our star-forming bulges are compatible with this scenario if these systems having undergone disk-bulge transformations, have shrunk, but have not yet quenched. However, while this basic model resulting in size shrinking followed by quenching, which by extension proposes that all quenched systems are equally compact, can, at least qualitatively, explain the existence and sizes of star-forming bulges, this is not true for passive disks, which have not undergone morphological or size transformations. This disk population can instead only be explained by secular quenching processes such e.g. halo quenching or ram pressure stripping, which do not transform the underlying morphology {\it or} sizes of the systems. This further highlights the necessity of models to account for the fact that compactness is here observed to correlate with {\it both} passivity {\it and} the presence of a bulge if they are to build a fully consistent scenario.

\section{Conclusions}

We have made use of the extended multiple-component bulge+disk morphological decomposition technique presented in Bruce et al. (2014a), which allowed us to conduct multiple-component SED fitting and has provided us with stellar-mass and star-formation rate estimates for the separate components. By combining these estimates with the decomposed morphological information we have explored the evolution of the most massive ($M_*>10^{11}\,{\rm M_{\odot}}$) galaxies at $1<z<3$ in terms of the trends witnessed in the size-mass relation and from the median sizes of these systems split by both morphology and star-formation activity.

Having conducted this analysis we have been able to examine the size-mass relations from the combined UDS and COSMOS samples and the relations which utilised the new decomposed stellar-mass estimates, and have found continued evidence that the bulge components display a stronger evolution in the size of the population compared to similarly massive local galaxies than the disk components. This can be seen from both the fraction of bulge components which lie below the local relation and the median sizes of the bulge components split above and below $z=2$. 

We have also found that, at $1<z<2$, $15\pm3\%$ of bulges have sizes consistent with the local ETG relation within its $1-\sigma$ scatter, with the median bulge component sizes being a factor of $2.93\pm0.32$ smaller than similarly massive local ETGs. At $2<z<3$ this fraction of bulges with sizes comparable to the local ETG relation becomes $23\pm6\%$ and median bulge component size is a factor of $3.41\pm0.58$ smaller than local ETGs. In comparison, at $1<z<2$, $49\pm6\%$ of the disk components have sizes consistent with the local LTG relation and its $1-\sigma$ scatter, and the median size is a factor of $1.65\pm0.14$ smaller. In the high-redshift bin these numbers become $43\pm8\%$ and the median size is a factor of $1.99\pm0.25$ smaller. The scatter in both the bulge and disk relations is larger than the measurement error and thus reflects the intrinsic scatter in the size-mass relations. However, by incorporating the new star-formation rate estimates from the decomposed SED fitting we do not find a clear distinction in the position of the passive and star-forming components on the size-mass relations. 

In order to further explore how the star-formation activity correlates with galaxy size we have examined the sizes of the galaxies in our sample as a fraction of the sizes of local similarly massive galaxies, split into the passive and star-forming bulge and disk sub-populations and plotted above and below $z=2$. By constructing these samples based on both the $n=2.5$ single-S\'{e}rsic fits with the overall star-formation rates, and using the decomposed morphologies and star-formation rates, we have highlighted the advantages in decomposing these galaxy properties. This analysis reveals that, while the single-S\'{e}rsic fits would indicate that the star-forming and passive bulges, and passive disks are equally compact in size (although the robustness of the star-forming bulge sample is questionable due to the high level of contamination from star-forming disk components; Bruce et al. 2014a) with the star-forming disks having larger sizes, the decomposed fits show that the passive disks have intermediate sizes. As the single-S\'{e}rsic fractional sizes are in broad agreement with results from previous studies, this clearly demonstrates that adopting the single-S\'{e}rsic fits presents a simplified view of the evolutionary processes involved, where for bulge-dominated systems morphology is the main
indicator of compactness and for disks the main indicator is star-formation. In comparison, the decomposed fits reveal that compactness correlates with some combination of passivity and the presence of a significant bulge component.

Moreover, by assuming that the star-forming disks are the direct progenitors of the passive disks, and by evolving the star-formation histories of the passive disks back to the redshifts $\sim1$\,Gyr earlier, when the passive disks were still active, we have shown that the passive disks and their star-forming progenitor disks have consistent sizes.

\section{Acknowledgments}
VAB acknowledges the support of the Science and Technology Facilities
Council (STFC) via the award of an STFC Studentship.
VAB and JSD acknowledge the support of the EC FP7 Space project ASTRODEEP (Ref. No: 312725).
JSD, RAAB, FB, TAT acknowledge the support of the 
European Research Council via the award of an Advanced Grant. 
JSD and RJM acknowledge the support 
of the Royal Society via a Wolfson Research Merit Award and a University 
Research Fellowship respectively.
RJM acknowledges the support of the Leverhulme Trust via the award of a Philip Leverhulme Research Prize.
MC acknowledges the support of the Science and Technology Facilities
Council (STFC) via the award of an  STFC Advanced 
Fellowship.
US authors acknowledge  support from NASA grants for {\it HST} Program GO-12060.
SMF, DCK, DDK, and EJM also acknowledge support from NSF grant AST-08-08133.

This work is based in part on observations made with the NASA/ESA {\it Hubble Space Telescope}, which is operated by the Association 
of Universities for Research in Astronomy, Inc, under NASA contract NAS5-26555.
This work is based in part on observations made with the {\it Spitzer Space Telescope}, which is operated by the Jet Propulsion Laboratory, 
California Institute of Technology under NASA contract 1407.

\nocite{Bruce2014a}
 
\bibliographystyle{mn2e}

\bibliography{bibtex}

\begin{thebibliography}{78}
\expandafter\ifx\csname natexlab\endcsname\relax\def\natexlab#1{#1}\fi

\bibitem[{{Allen} {et~al}\mbox{.}(2006){Allen}, {Driver}, {Graham}, {Cameron},
  {Liske}, \& {de Propris}}]{Allen2006}
{Allen} P.~D., {Driver} S.~P., {Graham} A.~W., {Cameron} E., {Liske} J., {de
  Propris} R., 2006, MNRAS, 371, 2

\bibitem[{{Barden} {et~al}\mbox{.}(2012){Barden}, {H{\"a}u{\ss}ler}, {Peng},
  {McIntosh}, \& {Guo}}]{Barden2012}
{Barden} M., {H{\"a}u{\ss}ler} B., {Peng} C.~Y., {McIntosh} D.~H., {Guo} Y.,
  2012, MNRAS, 422, 449

\bibitem[{{Barro} {et~al}\mbox{.}(2013{\natexlab{a}}){Barro}, {Faber},
  {P{\'e}rez-Gonz{\'a}lez}, {Koo}, {Williams}, {Kocevski}, {Trump}, {Mozena},
  {McGrath}, {van der Wel}, {Wuyts}, {Bell}, {Croton}, {Ceverino}, {Dekel},
  {Ashby}, {Cheung}, {Ferguson}, {Fontana}, {Fang}, {Giavalisco}, {Grogin},
  {Guo}, {Hathi}, {Hopkins}, {Huang}, {Koekemoer}, {Kartaltepe}, {Lee},
  {Newman}, {Porter}, {Primack}, {Ryan}, {Rosario}, {Somerville}, {Salvato}, \&
  {Hsu}}]{Barro2013}
{Barro} G. {et~al.}, 2013{\natexlab{a}}, ApJ, 765, 104

\bibitem[{{Barro} {et~al}\mbox{.}(2013{\natexlab{b}}){Barro}, {Faber},
  {Perez-Gonzalez}, {Pacifici}, {Trump}, {Koo}, {Wuyts}, {Guo}, {Bell},
  {Dekel}, {Porter}, {Primack}, {Ferguson}, {Ashby}, {Caputi}, {Ceverino},
  {Croton}, {Fazio}, {Giavalisco}, {Hsu}, {Kocevski}, {Koekemoer},
  {Kurczynski}, {Kollipara}, {Lee}, {McIntosh}, {McGrath}, {Moody},
  {Somerville}, {Papovich}, {Salvato}, {Santini}, {Williams}, {Willner}, \&
  {Zolotov}}]{Barro2013b}
{Barro} G. {et~al.}, 2013{\natexlab{b}}, ApJ, submitted (arXiv:1311.5559)

\bibitem[{{Belli} {et~al}\mbox{.}(2014){Belli}, {Newman}, \&
  {Ellis}}]{Belli2014}
{Belli} S., {Newman} A.~B., {Ellis} R.~S., 2014, ApJ, 783, 117

\bibitem[{{Bolzonella} {et~al}\mbox{.}(2000){Bolzonella}, {Miralles}, \&
  {Pell{\'o}}}]{Bolzonella2000}
{Bolzonella} M., {Miralles} J., {Pell{\'o}} R., 2000, A\&A, 363, 476

\bibitem[{{Bournaud} {et~al}\mbox{.}(2011){Bournaud}, {Dekel}, {Teyssier},
  {Cacciato}, {Daddi}, {Juneau}, \& {Shankar}}]{Bournaud2011}
{Bournaud} F., {Dekel} A., {Teyssier} R., {Cacciato} M., {Daddi} E., {Juneau}
  S., {Shankar} F., 2011, ApJL, 741, L33

\bibitem[{{Bruce} {et~al}\mbox{.}(2012){Bruce}, {Dunlop}, {Cirasuolo},
  {McLure}, {Targett}, {Bell}, {Croton}, {Dekel}, {Faber}, {Ferguson},
  {Grogin}, {Kocevski}, {Koekemoer}, {Koo}, {Lai}, {Lotz}, {McGrath}, {Newman},
  \& {van der Wel}}]{Bruce2012}
{Bruce} V.~A. {et~al.}, 2012, MNRAS, 427, 1666

\bibitem[{{Bruce} {et~al}\mbox{.}(2014){Bruce}, {Dunlop}, {McLure},
  {Cirasuolo}, {Buitrago}, {Bowler}, {Targett}, {Bell}, {McIntosh}, {Dekel},
  {Faber}, {Ferguson}, {Grogin}, {Hartley}, {Kocevski}, {Koekemoer}, {Koo}, \&
  {McGrath}}]{Bruce2014a}
{Bruce} V.~A. {et~al.}, 2014, MNRAS, submitted (arXiv:1405.1736)

\bibitem[{{Bruzual} \& {Charlot}(2003)}]{Bruzual2003}
{Bruzual} G., {Charlot} S., 2003, MNRAS, 344, 1000

\bibitem[{{Buitrago} {et~al}\mbox{.}(2008){Buitrago}, {Trujillo}, {Conselice},
  {Bouwens}, {Dickinson}, \& {Yan}}]{Buitrago2008}
{Buitrago} F., {Trujillo} I., {Conselice} C.~J., {Bouwens} R.~J., {Dickinson}
  M., {Yan} H., 2008, ApJL, 687, L61

\bibitem[{{Buitrago} {et~al}\mbox{.}(2013){Buitrago}, {Trujillo}, {Conselice},
  \& {H{\"a}u{\ss}ler}}]{Buitrago2011}
{Buitrago} F., {Trujillo} I., {Conselice} C.~J., {H{\"a}u{\ss}ler} B., 2013,
  MNRAS, 428, 1460

\bibitem[{{Cameron} {et~al}\mbox{.}(2009){Cameron}, {Driver}, {Graham}, \&
  {Liske}}]{Cameron2009}
{Cameron} E., {Driver} S.~P., {Graham} A.~W., {Liske} J., 2009, ApJ, 699, 105

\bibitem[{{Carollo} {et~al}\mbox{.}(2013){Carollo}, {Bschorr}, {Renzini},
  {Lilly}, {Capak}, {Cibinel}, {Ilbert}, {Onodera}, {Scoville}, {Cameron},
  {Mobasher}, {Sanders}, \& {Taniguchi}}]{Carollo2013}
{Carollo} C.~M. {et~al.}, 2013, ApJ, 773, 112

\bibitem[{{Cassata} {et~al}\mbox{.}(2010){Cassata}, {Giavalisco}, {Guo},
  {Ferguson}, {Koekemoer}, {Renzini}, {Fontana}, {Salimbeni}, {Dickinson},
  {Casertano}, {Conselice}, {Grogin}, {Lotz}, {Papovich}, {Lucas}, {Straughn},
  {Gardner}, \& {Moustakas}}]{Cassata2010}
{Cassata} P. {et~al.}, 2010, ApJL, 714, L79

\bibitem[{{Cassata} {et~al}\mbox{.}(2011){Cassata}, {Giavalisco}, {Guo},
  {Renzini}, {Ferguson}, {Koekemoer}, {Salimbeni}, {Scarlata}, {Grogin},
  {Conselice}, {Dahlen}, {Lotz}, {Dickinson}, \& {Lin}}]{Cassata2011}
{Cassata} P. {et~al.}, 2011, ApJ, 743, 96

\bibitem[{{Cassata} {et~al}\mbox{.}(2013){Cassata}, {Giavalisco}, {Williams},
  {Guo}, {Lee}, {Renzini}, {Ferguson}, {Faber}, {Barro}, {McIntosh}, {Lu},
  {Bell}, {Koo}, {Papovich}, {Ryan}, {Conselice}, {Grogin}, {Koekemoer}, \&
  {Hathi}}]{Cassata2013}
{Cassata} P. {et~al.}, 2013, ApJ, 775, 106

\bibitem[{{Cimatti} {et~al}\mbox{.}(2008){Cimatti}, {Cassata}, {Pozzetti},
  {Kurk}, {Mignoli}, {Renzini}, {Daddi}, {Bolzonella}, {Brusa}, {Rodighiero},
  {Dickinson}, {Franceschini}, {Zamorani}, {Berta}, {Rosati}, \&
  {Halliday}}]{Cimatti2008}
{Cimatti} A. {et~al.}, 2008, A\&A, 482, 21

\bibitem[{{Cirasuolo} {et~al}\mbox{.}(2007){Cirasuolo}, {McLure}, {Dunlop},
  {Almaini}, {Foucaud}, {Smail}, {Sekiguchi}, {Simpson}, {Eales}, {Dye},
  {Watson}, {Page}, \& {Hirst}}]{Cirasuolo2007}
{Cirasuolo} M. {et~al.}, 2007, MNRAS, 380, 585

\bibitem[{{Daddi} {et~al}\mbox{.}(2005){Daddi}, {Renzini}, {Pirzkal},
  {Cimatti}, {Malhotra}, {Stiavelli}, {Xu}, {Pasquali}, {Rhoads}, {Brusa}, {di
  Serego Alighieri}, {Ferguson}, {Koekemoer}, {Moustakas}, {Panagia}, \&
  {Windhorst}}]{Daddi2005}
{Daddi} E. {et~al.}, 2005, ApJ, 626, 680

\bibitem[{{Damjanov} {et~al}\mbox{.}(2009){Damjanov}, {McCarthy}, {Abraham},
  {Glazebrook}, {Yan}, {Mentuch}, {Le Borgne}, {Savaglio}, {Crampton},
  {Murowinski}, {Juneau}, {Carlberg}, {J{\o}rgensen}, {Roth}, {Chen}, \&
  {Marzke}}]{Damjanov2009}
{Damjanov} I. {et~al.}, 2009, ApJ, 695, 101

\bibitem[{{Davari} {et~al}\mbox{.}(2014){Davari}, {Ho}, {Peng}, \&
  {Huang}}]{Davari2014}
{Davari} R., {Ho} L.~C., {Peng} C.~Y., {Huang} S., 2014, ApJ, 787, 69

\bibitem[{{de Jong}(1996)}]{deJong1996}
{de Jong} R.~S., 1996, A\&A, 313, 45

\bibitem[{{Dekel} \& {Burkert}(2014)}]{Dekel2014}
{Dekel} A., {Burkert} A., 2014, MNRAS, 438, 1870

\bibitem[{{Fan} {et~al}\mbox{.}(2010){Fan}, {Lapi}, {Bressan}, {Bernardi}, {De
  Zotti}, \& {Danese}}]{Fan2010}
{Fan} L., {Lapi} A., {Bressan} A., {Bernardi} M., {De Zotti} G., {Danese} L.,
  2010, ApJ, 718, 1460

\bibitem[{{Fan} {et~al}\mbox{.}(2008){Fan}, {Lapi}, {De Zotti}, \&
  {Danese}}]{Fan2008}
{Fan} L., {Lapi} A., {De Zotti} G., {Danese} L., 2008, ApJL, 689, L101

\bibitem[{{Franx} {et~al}\mbox{.}(2008){Franx}, {van Dokkum}, {Schreiber},
  {Wuyts}, {Labb{\'e}}, \& {Toft}}]{Franx2008}
{Franx} M., {van Dokkum} P.~G., {Schreiber} N.~M.~F., {Wuyts} S., {Labb{\'e}}
  I., {Toft} S., 2008, ApJ, 688, 770

\bibitem[{{Furusawa} {et~al}\mbox{.}(2008){Furusawa}, {Kosugi}, {Akiyama},
  {Takata}, {Sekiguchi}, {Tanaka}, {Iwata}, {Kajisawa}, {Yasuda}, {Doi},
  {Ouchi}, {Simpson}, {Shimasaku}, {Yamada}, {Furusawa}, {Morokuma}, {Ishida},
  {Aoki}, {Fuse}, {Imanishi}, {Iye}, {Karoji}, {Kobayashi}, {Kodama},
  {Komiyama}, {Maeda}, {Miyazaki}, {Mizumoto}, {Nakata}, {Noumaru},
  {Ogasawara}, {Okamura}, {Saito}, {Sasaki}, {Ueda}, \&
  {Yoshida}}]{Furusawa2008}
{Furusawa} H. {et~al.}, 2008, ApJS, 176, 1

\bibitem[{{Grogin} {et~al}\mbox{.}(2011){Grogin}, {Kocevski}, {Faber},
  {Ferguson}, {Koekemoer}, {Riess}, {Acquaviva}, {Alexander}, {Almaini},
  {Ashby}, {Barden}, {Bell}, {Bournaud}, {Brown}, {Caputi}, {Casertano},
  {Cassata}, {Castellano}, {Challis}, {Chary}, {Cheung}, {Cirasuolo},
  {Conselice}, {Roshan Cooray}, {Croton}, {Daddi}, {Dahlen}, {Dav{\'e}}, {de
  Mello}, {Dekel}, {Dickinson}, {Dolch}, {Donley}, {Dunlop}, {Dutton}, {Elbaz},
  {Fazio}, {Filippenko}, {Finkelstein}, {Fontana}, {Gardner}, {Garnavich},
  {Gawiser}, {Giavalisco}, {Grazian}, {Guo}, {Hathi}, {H{\"a}ussler},
  {Hopkins}, {Huang}, {Huang}, {Jha}, {Kartaltepe}, {Kirshner}, {Koo}, {Lai},
  {Lee}, {Li}, {Lotz}, {Lucas}, {Madau}, {McCarthy}, {McGrath}, {McIntosh},
  {McLure}, {Mobasher}, {Moustakas}, {Mozena}, {Nandra}, {Newman}, {Niemi},
  {Noeske}, {Papovich}, {Pentericci}, {Pope}, {Primack}, {Rajan},
  {Ravindranath}, {Reddy}, {Renzini}, {Rix}, {Robaina}, {Rodney}, {Rosario},
  {Rosati}, {Salimbeni}, {Scarlata}, {Siana}, {Simard}, {Smidt}, {Somerville},
  {Spinrad}, {Straughn}, {Strolger}, {Telford}, {Teplitz}, {Trump}, {van der
  Wel}, {Villforth}, {Wechsler}, {Weiner}, {Wiklind}, {Wild}, {Wilson},
  {Wuyts}, {Yan}, \& {Yun}}]{Grogin2011}
{Grogin} N.~A. {et~al.}, 2011, ApJS, 197, 35

\bibitem[{{H\"{a}ussler} {et~al}\mbox{.}(2007){H\"{a}ussler}, {McIntosh},
  {Barden}, {Bell}, {Rix}, {Borch}, {Beckwith}, {Caldwell}, {Heymans},
  {Jahnke}, {Jogee}, {Koposov}, {Meisenheimer}, {S{\'a}nchez}, {Somerville},
  {Wisotzki}, \& {Wolf}}]{Haussler2007}
{H\"{a}ussler} B. {et~al.}, 2007, ApJS, 172, 615

\bibitem[{{Hopkins} {et~al}\mbox{.}(2009){Hopkins}, {Cox}, {Younger}, \&
  {Hernquist}}]{Hopkins2009}
{Hopkins} P.~F., {Cox} T.~J., {Younger} J.~D., {Hernquist} L., 2009, ApJ, 691,
  1168

\bibitem[{{Kennicutt}(1998)}]{Kennicutt1998}
{Kennicutt}, Jr. R.~C., 1998, ARA\&A, 36, 189

\bibitem[{{Khochfar} \& {Silk}(2006)}]{Khochfar2006}
{Khochfar} S., {Silk} J., 2006, ApJL, 648, L21

\bibitem[{{Koekemoer} {et~al}\mbox{.}(2007){Koekemoer}, {Aussel}, {Calzetti},
  {Capak}, {Giavalisco}, {Kneib}, {Leauthaud}, {Le F{\`e}vre}, {McCracken},
  {Massey}, {Mobasher}, {Rhodes}, {Scoville}, \& {Shopbell}}]{Koekemoer2007}
{Koekemoer} A.~M. {et~al.}, 2007, ApJS, 172, 196

\bibitem[{{Koekemoer} {et~al}\mbox{.}(2011){Koekemoer}, {Faber}, {Ferguson},
  {Grogin}, {Kocevski}, {Koo}, {Lai}, {Lotz}, {Lucas}, {McGrath}, {Ogaz},
  {Rajan}, {Riess}, {Rodney}, {Strolger}, {Casertano}, {Castellano}, {Dahlen},
  {Dickinson}, {Dolch}, {Fontana}, {Giavalisco}, {Grazian}, {Guo}, {Hathi},
  {Huang}, {van der Wel}, {Yan}, {Acquaviva}, {Alexander}, {Almaini}, {Ashby},
  {Barden}, {Bell}, {Bournaud}, {Brown}, {Caputi}, {Cassata}, {Challis},
  {Chary}, {Cheung}, {Cirasuolo}, {Conselice}, {Roshan Cooray}, {Croton},
  {Daddi}, {Dav{\'e}}, {de Mello}, {de Ravel}, {Dekel}, {Donley}, {Dunlop},
  {Dutton}, {Elbaz}, {Fazio}, {Filippenko}, {Finkelstein}, {Frazer}, {Gardner},
  {Garnavich}, {Gawiser}, {Gruetzbauch}, {Hartley}, {H{\"a}ussler},
  {Herrington}, {Hopkins}, {Huang}, {Jha}, {Johnson}, {Kartaltepe},
  {Khostovan}, {Kirshner}, {Lani}, {Lee}, {Li}, {Madau}, {McCarthy},
  {McIntosh}, {McLure}, {McPartland}, {Mobasher}, {Moreira}, {Mortlock},
  {Moustakas}, {Mozena}, {Nandra}, {Newman}, {Nielsen}, {Niemi}, {Noeske},
  {Papovich}, {Pentericci}, {Pope}, {Primack}, {Ravindranath}, {Reddy},
  {Renzini}, {Rix}, {Robaina}, {Rosario}, {Rosati}, {Salimbeni}, {Scarlata},
  {Siana}, {Simard}, {Smidt}, {Snyder}, {Somerville}, {Spinrad}, {Straughn},
  {Telford}, {Teplitz}, {Trump}, {Vargas}, {Villforth}, {Wagner}, {Wandro},
  {Wechsler}, {Weiner}, {Wiklind}, {Wild}, {Wilson}, {Wuyts}, \&
  {Yun}}]{Koekemoer2011}
{Koekemoer} A.~M. {et~al.}, 2011, ApJS, 197, 36

\bibitem[{{Kriek} {et~al}\mbox{.}(2009){Kriek}, {van Dokkum}, {Franx},
  {Illingworth}, \& {Magee}}]{Kriek2009}
{Kriek} M., {van Dokkum} P.~G., {Franx} M., {Illingworth} G.~D., {Magee} D.~K.,
  2009, ApJL, 705, L71

\bibitem[{{Kriek} {et~al}\mbox{.}(2006){Kriek}, {van Dokkum}, {Franx},
  {Quadri}, {Gawiser}, {Herrera}, {Illingworth}, {Labb{\'e}}, {Lira},
  {Marchesini}, {Rix}, {Rudnick}, {Taylor}, {Toft}, {Urry}, \&
  {Wuyts}}]{Kriek2006}
{Kriek} M. {et~al.}, 2006, ApJL, 649, L71

\bibitem[{{Krogager} {et~al}\mbox{.}(2013){Krogager}, {Zirm}, {Toft}, {Man}, \&
  {Brammer}}]{Krogager2013}
{Krogager} J.-K., {Zirm} A.~W., {Toft} S., {Man} A., {Brammer} G., 2013, ApJ,
  submitted (arXix:1309.6316)

\bibitem[{{Lackner} \& {Gunn}(2012)}]{Lackner2012}
{Lackner} C.~N., {Gunn} J.~E., 2012, MNRAS, 421, 2277

\bibitem[{{Lang} {et~al}\mbox{.}(2014){Lang}, {Wuyts}, {Somerville}, {Forster
  Schreiber}, {Genzel}, {Bell}, {Brammer}, {Dekel}, {Faber}, {Ferguson},
  {Grogin}, {Kocevski}, {Koekemoer}, {Lutz}, {McGrath}, {Momcheva}, {Nelson},
  {Primack}, {Rosario}, {Skelton}, {Tacconi}, {van Dokkum}, \&
  {Whitaker}}]{Lang2014}
{Lang} P. {et~al.}, 2014, ApJ, accepted (arXiv:1402.0866)

\bibitem[{{Lawrence} {et~al}\mbox{.}(2007){Lawrence}, {Warren}, {Almaini},
  {Edge}, {Hambly}, {Jameson}, {Lucas}, {Casali}, {Adamson}, {Dye}, {Emerson},
  {Foucaud}, {Hewett}, {Hirst}, {Hodgkin}, {Irwin}, {Lodieu}, {McMahon},
  {Simpson}, {Smail}, {Mortlock}, \& {Folger}}]{Lawrence2007}
{Lawrence} A. {et~al.}, 2007, MNRAS, 379, 1599

\bibitem[{{Mancini} {et~al}\mbox{.}(2010){Mancini}, {Daddi}, {Renzini},
  {Salmi}, {McCracken}, {Cimatti}, {Onodera}, {Salvato}, {Koekemoer}, {Aussel},
  {Le Floc'h}, {Willott}, \& {Capak}}]{Mancini2010}
{Mancini} C. {et~al.}, 2010, MNRAS, 401, 933

\bibitem[{{McLure} {et~al}\mbox{.}(2013){McLure}, {Pearce}, {Dunlop},
  {Cirasuolo}, {Curtis-Lake}, {Bruce}, {Caputi}, {Almaini}, {Bonfield},
  {Bradshaw}, {Buitrago}, {Chuter}, {Foucaud}, {Hartley}, \&
  {Jarvis}}]{McLure2013}
{McLure} R.~J. {et~al.}, 2013, MNRAS, 428, 1088

\bibitem[{{Mortlock} {et~al}\mbox{.}(2013){Mortlock}, {Conselice}, {Hartley},
  {Ownsworth}, {Lani}, {Bluck}, {Almaini}, {Duncan}, {Wel}, {Koekemoer},
  {Dekel}, {Dav{\'e}}, {Ferguson}, {de Mello}, {Newman}, {Faber}, {Grogin},
  {Kocevski}, \& {Lai}}]{Mortlock2013}
{Mortlock} A. {et~al.}, 2013, MNRAS, 433, 1185

\bibitem[{{Mozena} {et~al}\mbox{.}(2013){Mozena}, {Faber}, {Koo}, {Primack},
  {Dekel}, {Moody}, {Ceverino}, \& {CANDELS}}]{Mozena2013}
{Mozena} M., {Faber} S.~M., {Koo} D.~C., {Primack} J.~R., {Dekel} A., {Moody}
  C.~E., {Ceverino} D., {CANDELS}, 2013, in American Astronomical Society
  Meeting Abstracts, Vol. 221, American Astronomical Society Meeting Abstracts,
  p. 112

\bibitem[{{Muzzin} {et~al}\mbox{.}(2009){Muzzin}, {van Dokkum}, {Franx},
  {Marchesini}, {Kriek}, \& {Labb{\'e}}}]{Muzzin2009}
{Muzzin} A., {van Dokkum} P., {Franx} M., {Marchesini} D., {Kriek} M.,
  {Labb{\'e}} I., 2009, ApJL, 706, L188

\bibitem[{{Naab} {et~al}\mbox{.}(2007){Naab}, {Johansson}, {Ostriker}, \&
  {Efstathiou}}]{Naab2007}
{Naab} T., {Johansson} P.~H., {Ostriker} J.~P., {Efstathiou} G., 2007, ApJ,
  658, 710

\bibitem[{{Newman} {et~al}\mbox{.}(2012){Newman}, {Ellis}, {Bundy}, \&
  {Treu}}]{Newman2012}
{Newman} A.~B., {Ellis} R.~S., {Bundy} K., {Treu} T., 2012, ApJ, 746, 162

\bibitem[{{Newman} {et~al}\mbox{.}(2010){Newman}, {Ellis}, {Treu}, \&
  {Bundy}}]{Newman2010}
{Newman} A.~B., {Ellis} R.~S., {Treu} T., {Bundy} K., 2010, ApJL, 717, L103

\bibitem[{{Peng} {et~al}\mbox{.}(2010){Peng}, {Ho}, {Impey}, \&
  {Rix}}]{Peng2010galfit}
{Peng} C.~Y., {Ho} L.~C., {Impey} C.~D., {Rix} H.-W., 2010, AJ, 139, 2097

\bibitem[{{Poggianti} {et~al}\mbox{.}(2013{\natexlab{a}}){Poggianti}, {Calvi},
  {Bindoni}, {D'Onofrio}, {Moretti}, {Valentinuzzi}, {Fasano}, {Fritz}, {De
  Lucia}, {Vulcani}, {Bettoni}, {Gullieuszik}, \& {Omizzolo}}]{Poggianti2013}
{Poggianti} B.~M. {et~al.}, 2013{\natexlab{a}}, ApJ, 762, 77

\bibitem[{{Poggianti} {et~al}\mbox{.}(2013{\natexlab{b}}){Poggianti},
  {Moretti}, {Calvi}, {D'Onofrio}, {Valentinuzzi}, {Fritz}, \&
  {Renzini}}]{Poggianti2013b}
{Poggianti} B.~M., {Moretti} A., {Calvi} R., {D'Onofrio} M., {Valentinuzzi} T.,
  {Fritz} J., {Renzini} A., 2013{\natexlab{b}}, ApJ, 777, 125

\bibitem[{{Scoville} {et~al}\mbox{.}(2007){Scoville}, {Aussel}, {Brusa},
  {Capak}, {Carollo}, {Elvis}, {Giavalisco}, {Guzzo}, {Hasinger}, {Impey},
  {Kneib}, {LeFevre}, {Lilly}, {Mobasher}, {Renzini}, {Rich}, {Sanders},
  {Schinnerer}, {Schminovich}, {Shopbell}, {Taniguchi}, \&
  {Tyson}}]{Scoville2007}
{Scoville} N. {et~al.}, 2007, ApJS, 172, 1

\bibitem[{{Sekiguchi} {et~al}\mbox{.}(2005){Sekiguchi}, {Akiyama}, {Furusawa},
  {Simpson}, {Takata}, {Ueda}, {Watson}, \& {Sxds Team}}]{Sekiguchi2005}
{Sekiguchi} K., {Akiyama} M., {Furusawa} H., {Simpson} C., {Takata} T., {Ueda}
  Y., {Watson} M.~W., {Sxds Team}, 2005, in Multiwavelength Mapping of Galaxy
  Formation and Evolution, {A.~Renzini \& R.~Bender}, ed., p.~82

\bibitem[{{Shankar} {et~al}\mbox{.}(2011){Shankar}, {Marulli}, {Bernardi},
  {Mei}, {Meert}, \& {Vikram}}]{Shankar2011}
{Shankar} F., {Marulli} F., {Bernardi} M., {Mei} S., {Meert} A., {Vikram} V.,
  2011, MNRAS, submitted (arXiv:1105.6043)

\bibitem[{{Shen} {et~al}\mbox{.}(2003){Shen}, {Mo}, {White}, {Blanton},
  {Kauffmann}, {Voges}, {Brinkmann}, \& {Csabai}}]{Shen2003}
{Shen} S., {Mo} H.~J., {White} S.~D.~M., {Blanton} M.~R., {Kauffmann} G.,
  {Voges} W., {Brinkmann} J., {Csabai} I., 2003, MNRAS, 343, 978

\bibitem[{{Simard} {et~al}\mbox{.}(2011){Simard}, {Mendel}, {Patton},
  {Ellison}, \& {McConnachie}}]{Simard2011}
{Simard} L., {Mendel} J.~T., {Patton} D.~R., {Ellison} S.~L., {McConnachie}
  A.~W., 2011, ApJS, 196, 11

\bibitem[{{Szomoru} {et~al}\mbox{.}(2012){Szomoru}, {Franx}, \& {van
  Dokkum}}]{Szomoru2012}
{Szomoru} D., {Franx} M., {van Dokkum} P.~G., 2012, ApJ, 749, 121

\bibitem[{{Szomoru} {et~al}\mbox{.}(2010){Szomoru}, {Franx}, {van Dokkum},
  {Trenti}, {Illingworth}, {Labb{\'e}}, {Bouwens}, {Oesch}, \&
  {Carollo}}]{Szomoru2010}
{Szomoru} D. {et~al.}, 2010, ApJL, 714, L244

\bibitem[{{Taylor} {et~al}\mbox{.}(2010){Taylor}, {Franx}, {Glazebrook},
  {Brinchmann}, {van der Wel}, \& {van Dokkum}}]{Taylor2010}
{Taylor} E.~N., {Franx} M., {Glazebrook} K., {Brinchmann} J., {van der Wel} A.,
  {van Dokkum} P.~G., 2010, ApJ, 720, 723

\bibitem[{{Toft} {et~al}\mbox{.}(2007){Toft}, {van Dokkum}, {Franx}, {Labbe},
  {F{\"o}rster Schreiber}, {Wuyts}, {Webb}, {Rudnick}, {Zirm}, {Kriek}, {van
  der Werf}, {Blakeslee}, {Illingworth}, {Rix}, {Papovich}, \&
  {Moorwood}}]{Toft2007}
{Toft} S. {et~al.}, 2007, ApJ, 671, 285

\bibitem[{{Trujillo} {et~al}\mbox{.}(2012){Trujillo}, {Carrasco}, \&
  {Ferr{\'e}-Mateu}}]{Trujillo2012}
{Trujillo} I., {Carrasco} E.~R., {Ferr{\'e}-Mateu} A., 2012, ApJ, 751, 45

\bibitem[{{Trujillo} {et~al}\mbox{.}(2009){Trujillo}, {Cenarro}, {de
  Lorenzo-C{\'a}ceres}, {Vazdekis}, {de la Rosa}, \& {Cava}}]{Trujillo2009}
{Trujillo} I., {Cenarro} A.~J., {de Lorenzo-C{\'a}ceres} A., {Vazdekis} A., {de
  la Rosa} I.~G., {Cava} A., 2009, ApJL, 692, L118

\bibitem[{{Trujillo} {et~al}\mbox{.}(2007){Trujillo}, {Conselice}, {Bundy},
  {Cooper}, {Eisenhardt}, \& {Ellis}}]{Trujillo2007}
{Trujillo} I., {Conselice} C.~J., {Bundy} K., {Cooper} M.~C., {Eisenhardt} P.,
  {Ellis} R.~S., 2007, MNRAS, 382, 109

\bibitem[{{Trujillo} {et~al}\mbox{.}(2006){Trujillo}, {F{\"o}rster Schreiber},
  {Rudnick}, {Barden}, {Franx}, {Rix}, {Caldwell}, {McIntosh}, {Toft},
  {H{\"a}ussler}, {Zirm}, {van Dokkum}, {Labb{\'e}}, {Moorwood},
  {R{\"o}ttgering}, {van der Wel}, {van der Werf}, \& {van
  Starkenburg}}]{Trujillo2006}
{Trujillo} I. {et~al.}, 2006, ApJ, 650, 18

\bibitem[{{Valentinuzzi} {et~al}\mbox{.}(2010{\natexlab{a}}){Valentinuzzi},
  {Fritz}, {Poggianti}, {Cava}, {Bettoni}, {Fasano}, {D'Onofrio}, {Couch},
  {Dressler}, {Moles}, {Moretti}, {Omizzolo}, {Kj{\ae}rgaard}, {Vanzella}, \&
  {Varela}}]{Valentinuzzi2010a}
{Valentinuzzi} T. {et~al.}, 2010{\natexlab{a}}, ApJ, 712, 226

\bibitem[{{Valentinuzzi} {et~al}\mbox{.}(2010{\natexlab{b}}){Valentinuzzi},
  {Poggianti}, {Saglia}, {Arag{\'o}n-Salamanca}, {Simard},
  {S{\'a}nchez-Bl{\'a}zquez}, {D'onofrio}, {Cava}, {Couch}, {Fritz}, {Moretti},
  \& {Vulcani}}]{Valentinuzzi2010b}
{Valentinuzzi} T. {et~al.}, 2010{\natexlab{b}}, ApJL, 721, L19

\bibitem[{{van de Sande} {et~al}\mbox{.}(2013){van de Sande}, {Kriek}, {Franx},
  {van Dokkum}, {Bezanson}, {Bouwens}, {Quadri}, {Rix}, \&
  {Skelton}}]{vandeSande2013}
{van de Sande} J. {et~al.}, 2013, ApJ, 771, 85

\bibitem[{{van de Sande} {et~al}\mbox{.}(2011){van de Sande}, {Kriek}, {Franx},
  {van Dokkum}, {Bezanson}, {Whitaker}, {Brammer}, {Labb{\'e}}, {Groot}, \&
  {Kaper}}]{vandeSande2011}
{van de Sande} J. {et~al.}, 2011, ApJL, 736, L9

\bibitem[{{van der Wel} {et~al}\mbox{.}(2012){van der Wel}, {Bell},
  {H{\"a}ussler}, {McGrath}, {Chang}, {Guo}, {McIntosh}, {Rix}, {Barden},
  {Cheung}, {Faber}, {Ferguson}, {Galametz}, {Grogin}, {Hartley}, {Kartaltepe},
  {Kocevski}, {Koekemoer}, {Lotz}, {Mozena}, {Peth}, \& {Peng}}]{vanderWel2012}
{van der Wel} A. {et~al.}, 2012, ApJS, 203, 24

\bibitem[{{van der Wel} {et~al}\mbox{.}(2009){van der Wel}, {Bell}, {van den
  Bosch}, {Gallazzi}, \& {Rix}}]{Vanderwel2009}
{van der Wel} A., {Bell} E.~F., {van den Bosch} F.~C., {Gallazzi} A., {Rix}
  H.-W., 2009, ApJ, 698, 1232

\bibitem[{{van der Wel} {et~al}\mbox{.}(2014){van der Wel}, {Franx}, {van
  Dokkum}, {Skelton}, {Momcheva}, {Whitaker}, {Brammer}, {Bell}, {Rix},
  {Wuyts}, {Ferguson}, {Holden}, {Barro}, {Koekemoer}, {Chang}, {McGrath},
  {Haussler}, {Dekel}, {Behroozi}, {Fumagalli}, {Leja}, {Lundgren}, {Maseda},
  {Nelson}, {Wake}, {Patel}, {Labbe}, {Faber}, {Grogin}, \&
  {Kocevski}}]{vanderWel2014}
{van der Wel} A. {et~al.}, 2014, ApJ, in press (arXiv:1404.2844)

\bibitem[{{van der Wel} {et~al}\mbox{.}(2008){van der Wel}, {Holden}, {Zirm},
  {Franx}, {Rettura}, {Illingworth}, \& {Ford}}]{Vanderwel2008}
{van der Wel} A., {Holden} B.~P., {Zirm} A.~W., {Franx} M., {Rettura} A.,
  {Illingworth} G.~D., {Ford} H.~C., 2008, ApJ, 688, 48

\bibitem[{{van der Wel} {et~al}\mbox{.}(2011){van der Wel}, {Rix}, {Wuyts},
  {McGrath}, {Koekemoer}, {Bell}, {Holden}, {Robaina}, \&
  {McIntosh}}]{Vanderwel2011}
{van der Wel} A. {et~al.}, 2011, ApJ, 730, 38

\bibitem[{{van Dokkum} {et~al}\mbox{.}(2008){van Dokkum}, {Franx}, {Kriek},
  {Holden}, {Illingworth}, {Magee}, {Bouwens}, {Marchesini}, {Quadri},
  {Rudnick}, {Taylor}, \& {Toft}}]{vanDokkum2008}
{van Dokkum} P.~G. {et~al.}, 2008, ApJL, 677, L5

\bibitem[{{Williams} {et~al}\mbox{.}(2009){Williams}, {Quadri}, {Franx}, {van
  Dokkum}, \& {Labb{\'e}}}]{Williams2009}
{Williams} R.~J., {Quadri} R.~F., {Franx} M., {van Dokkum} P., {Labb{\'e}} I.,
  2009, ApJ, 691, 1879

\bibitem[{{Wuyts} {et~al}\mbox{.}(2012){Wuyts}, {F{\"o}rster Schreiber},
  {Genzel}, {Guo}, {Barro}, {Bell}, {Dekel}, {Faber}, {Ferguson}, {Giavalisco},
  {Grogin}, {Hathi}, {Huang}, {Kocevski}, {Koekemoer}, {Koo}, {Lotz}, {Lutz},
  {McGrath}, {Newman}, {Rosario}, {Saintonge}, {Tacconi}, {Weiner}, \& {van der
  Wel}}]{Wuyts2012}
{Wuyts} S. {et~al.}, 2012, ApJ, 753, 114

\bibitem[{{Wuyts} {et~al}\mbox{.}(2011){Wuyts}, {F{\"o}rster Schreiber},
  {Lutz}, {Nordon}, {Berta}, {Altieri}, {Andreani}, {Aussel}, {Bongiovanni},
  {Cepa}, {Cimatti}, {Daddi}, {Elbaz}, {Genzel}, {Koekemoer}, {Magnelli},
  {Maiolino}, {McGrath}, {P{\'e}rez Garc{\'{\i}}a}, {Poglitsch}, {Popesso},
  {Pozzi}, {Sanchez-Portal}, {Sturm}, {Tacconi}, \& {Valtchanov}}]{Wuyts2011}
{Wuyts} S. {et~al.}, 2011, ApJ, 738, 106

\end{thebibliography}

\label{lastpage}

\end{document}